\documentclass[reprint,amsmath,amssymb,aps,superscriptaddress,nofootinbib,twocolumn,10pt]{revtex4-1}
\usepackage[colorlinks,linkcolor=blue,citecolor=blue,urlcolor=blue]{hyperref}
\usepackage{graphicx,epstopdf}
\usepackage{dcolumn}
\usepackage{subfigure}
\usepackage{bm}
\usepackage{amsmath,amsfonts,amssymb,graphicx,multirow,dcolumn,bm,latexsym,soul,nicefrac}
\usepackage{acronym}
\usepackage{enumitem}
\usepackage{array}
\usepackage{multirow}
\usepackage{appendix}
\usepackage{longtable}
\usepackage{footnote}
\usepackage{amssymb}
\usepackage{ctex}
\newacro{GR}{general relativity}
\newacro{GW}{gravitational wave}
\newacro{MBH}{massive black hole}
\newacro{MBHB}{massive black hole binary}
\newacro{EMRI}{extreme mass-ratio inspiral}
\newacro{IMRI}{intermediate mass-ratio inspiral}
\newacro{ppE}{parameterized post-Einsteinian}
\newacro{BH}{black hole}
\newacro{NS}{neutron star}
\newacro{BS}{boson star}
\newacro{ECO}{exotic compact object}
\newacro{LVK}{LIGO-Virgo-KAGRA}
\newacro{BBH}{binary black hole}
\newacro{BNS}{binary neutron star}
\newacro{NSBH}{neutron star-black hole}
\newacro{SIQM}{spin induced quadrupole moment}
\newacro{ASD}{amplitude spectral density}
\newacro{PSD}{power spectral density}
\newacro{SNR}{signal-to-noise ratio}
\newacro{PE}{parameter estimation}
\newacro{PN}{post-Newtonian}
\newacro{FIM}{fisher information matrix}
\newacro{TDI}{time delay interferometry}
\newacro{ISCO}{innermost stable circular orbit}
\newacro{PDF}{probability distribution function}
\newacro{EPS}{extended Press and Schechter}
\newacro{LISA}{Laser Interferometer Space Antenna}
\newacro{ppE}{parameterized post-Einstein}
\newacro{DF}{dynamical friction}
\newacro{DM}{dark matter}
\newacro{IMRI}{intermediate mass-ratio inspiral}
\newacro{Df}{dynamical friction}
\newacro{DM}{dark matter}
\newacro{MBH}{massive black hole}
\newacro{RS}{Randall and Sundrum}
\newacro{EdGB}{Einstein-dilaton-Gauss-Bonnet}
\newacro{EA}{Einstein-Æther }
\newacro{ROC}{Receiver Operating Characteristic Curve}
\newacro{AUC}{area under curve}
\newacro{KBR}{Kerr-Bertotti-Robinson}
\newacro{KBM}{Kerr-Bonnor-Melvin}
\newacro{BR}{Bertotti-Robinson}
\newacro{BM}{Bonnor-Melvin}

\newcommand{\tx}[1]{\text{#1}}
\newcommand{\scf}[1]{\times 10^{#1}}
\newcommand{\td}[1]{\tilde{#1}}
\newcommand{\mc}[1]{\mathcal{#1}}

\newcommand{\figref}[1]{Fig.~\ref{#1}}

\renewcommand{\eqref}[1]{Eq. (\ref{#1})}

\def\mrd{\mathrm{d} }

\begin{document}

\title{Distinguishability of magnetic massive black holes from environmental mimics with inspiral gravitational waves}



\newcommand{\SPA}{School of Physics and Astronomy, Sun Yat-sen University (Zhuhai Campus),\\ Zhuhai 519082, China.}

\newcommand{\SCUTP}{School of Physics and Optoelectronics, South China University of Technology,\\Guangzhou 510641, China}
\author{Xulong Yuan (袁旭龙)}
\affiliation{\SPA}
\author{Xiangdong Zhang (张向东)}
\email{Corresponding author: scxdzhang@scut.edu.cn}
\affiliation{\SCUTP}




\begin{abstract}
	Magnetic fields represent a critical component of astrophysical research, laying the foundation for interpreting high-energy astrophysical activity across galactic scales. In this work, we investigate the parametrized post-Einsteinian (ppE) waveform imprints induced by the external magnetic fields of Bertotti-Robinson and Bonnor-Melvin black holes, with the aim of distinguishing such magnetic effects from environmental influences—particularly for massive black holes posited to reside at galactic centers. We first compute the ppE frequency-domain waveform for a small black hole inspiraling into a massive Kerr-Bertotti-Robinson (KBR) black hole, which corresponds to a Kerr black hole embedded in an external magnetic field. We find that the leading-order correction arising from the magnetic field is at the $-2$ post-Newtonian (PN) order relative to the quadrupole term, while the next-leading-order correction is at $-1.5$ PN, originating from the spin of the black hole. We further examine the effects of a spinning Kerr-Bonnor-Melvin (KBM) black hole, whose leading-order magnetic correction is at $-3$ PN (consistent with the preceding result), whereas its spin-induced correction is also at $-1.5$ PN. The leading-order ppE corrections for both KBR and KBM black holes do not appear degenerate with any modified theory of gravity effects; nonetheless, we demonstrate that they resemble the gravitational pull contributions from additional matter with power-law distributions of index $\gamma=1$ and $\gamma=0$, respectively. To break the degeneracy with a single event, we adopt the statistic $F$ in former research to discriminate between these two classes of beyond-vacuum general relativity (GR) effects using multiple gravitational wave events. Our result shows even with multiple event statistic, it is not always efficient to distinguish real magnetic field effect from corresponding gravitational pull effect, which arises when they are not strong enough and distributed highly similarly, especially for Bertotti-Robinson magnetic effect. For Bonnor-Melvin black hole, there is a transition value of $\rho_0$ estimated around $10^{-4}\text{kg}/\text{m}^3$ and corresponding $B\sim 10^{4}\text{T}$ above which real magnetic effect can be efficiently distinguished from gravitational pull and below the transition value it cannot. As a result, future gravitational wave (GW) observations detecting $-3$ or $-2$ PN order corrections may infer their origin as either magnetic field effects or matter environmental influences, so multiple event statistic and multi-messenger validation are both important. 
\end{abstract}

\maketitle

\section{Introduction}
\label{intro}


The detection of \ac{GW} from binary black hole merger has opened a new era of astronomy \cite{figw}. One of the most fascinating sources is massive black holes assumed to reside at the galactic centers, which may form bound system with various objects and emit \acp{GW}. Thus upcoming space borne gravitational wave detectors like LISA \cite{Danzmann:1997hm, LISA:2017pwj}, Tianqin \cite{Luo:2015ght,TianQin:2020hid,Luo:2025sos} and Taiji \cite{Hu:2017mde} are expected to probe them in the near future. The coalescence of \ac{MBHB} lasts for a long duration with numerous cycles of motion and \ac{GW} phase, which can shed light on the complex environment of the galactic center when the binary is influenced by beyond-vacuum \ac{GR} effects. Besides usually discussed environments such as accretion disks, \ac{DM} and third bodies \cite{Camilloni:2023xvf,Tahelyani:2024cvk,AbhishekChowdhuri:2023rfv}, magnetic fields are another important one. Observations show that magnetic fields are commonly present on large scale in galactic disks, halos \cite{astra-5-43-2009,Beck:2015jta} and galaxy clusters \cite{annurev:/content/journals/10.1146/annurev.astro.40.060401.093852,Osinga:2022tos,Hu:2023hve}. Similar to modified gravity theory effects, these factors may be mistaken for deviations from \ac{GR}, so it is important to distinguish between environmental effects and modified theories of gravity. 

Within the \ac{ppE} framework \cite{bppe,ppewv}, the binary inspiral waveform corrections arising from beyond-vacuum \ac{GR} effects can be expressed in parametric form. Waveform corrections for a variety of modified theories of gravity \cite{PhysRevD.65.042002,esbgtl,gdxw,PhysRevD.83.084036,PhysRevD.85.064022,PhysRevLett.109.251105,Hansen:2014ewa,PhysRevD.94.084002,Kobakhidze:2016cqh,PhysRevD.94.084002,Liu:2024lda,Liu:2024qci,Bravo-Gaete:2026com} have been computed for phase contributions ranging from $-4$ to $4$ PN order \cite{ppewv,tgrgb}. On the other hand, environments surrounding the compact binary also imprint signatures on the orbital evolution and the emitted \ac{GW} signal. To date, numerous studies have investigated their influence on orbital dynamics and waveform modifications \cite{kocsis,yunes,eda,pmm,Barausse:2014tra,constr,Zi:2026zpw,Ghosh:2025ban}, parameter estimation \cite{pmm,constr,dwavf,mdme,CanevaSantoro:2023aol,yuan2024,Yuan:2025pbu,Zi:2025onl} and model identification \cite{Cole:2022yzw,pracc,Kejriwal:2025jao}. 

Since environmental effects are dominant at large separations, the corresponding waveform corrections typically appear at negative PN orders, most notably at $-4$ PN and $-1$ PN, which characterize mass-varying effects and dipole radiation, respectively. In particular, corrections between $-4$ and $-1$ PN order are absent in modified gravity scenarios, but can arise naturally in environmental contexts--for example, from gravitational interactions or accretion onto matter distributions with certain power-law profiles, or from magnetic field effects   \cite{Barausse:2014tra,fclg,coaenv}. Previous analyses have shown that the \ac{ppE} waveform correction induced by an external magnetic field around a Schwarzschild–Bonnor–Melvin black hole \cite{Ernst:1976mzr} at $-3$ PN order \cite{Barausse:2014tra}. 

Thus, if a $-3$ PN order waveform correction is detected in future observations, it is unlikely to originate from modified gravity. Instead, within standard \ac{GR}, such a correction can arise if one of the black holes is endowed with an external magnetic field, which modifies the orbital energy balance and thereby imprints a -3 PN correction on the \ac{GW} signal. Such a signature could also stem from other environmental effects. It is therefore crucial to distinguish such non-vacuum \ac{GR} effects from genuine environmental contributions when testing \ac{GR}.

A spinning Bonnor–Melvin black hole corresponds to a \ac{KBM} black hole \cite{Ernst:1976bsr}. Interestingly, a new Kerr black hole solution with an external magnetic field--the \ac{KBR} black hole  \cite{Podolsky:2025tle} has recently been discovered, which avoids several drawbacks of the Bonnor–Melvin spacetime, such as geodesics that cannot escape to infinity and chaotic geodesic motion. Time-domain corrections for such spacetimes have recently been explored \cite{Li:2025rtf}. Thermodynamics of this \ac{KBR} blackhole is given in \cite{Hu:2026slp}. 

In this work, we derive the \ac{ppE} inspiral waveform correction for a binary system containing one \ac{KBR} black hole. We find that it produces a phase correction at $-2$ PN order, which again lies in the $-4$ to $-1$ PN range inaccessible to modified gravity theories. Given that the Bonnor–Melvin black hole yields a  $-3$ PN correction \cite{Barausse:2014tra}, we also include spin effects to obtain the waveform for binaries involving a \ac{KBM} black hole. Using these waveform corrections for \ac{KBR}- and \ac{KBM}-containing binaries, we analyze the prospects for detecting these two types of external magnetic fields with inspiral \ac{GW} observations using the TianQin detector.

\textcolor{black}{The second point of this paper is that the leading order waveform correction of the two magnetic fields can be degenerate with those of environmental effect: the gravitational pull of power-law matter distribution with some index $\gamma$. We will give the relation between parameters of two kinds of effects which generate equal waveform correction, thus \ac{GW} corrections of them in future observations might be mistaken as each other, especially in a single event.} In order to discriminate these \ac{GR}-consistent magnetic field effects from environmental influences, we will adopt the statistic $F$ and methodology developed by Yuan et al. \cite{yuan2024,Yuan:2025pbu} to distinguish \ac{KBR} and \ac{KBM} magnetic field signatures from environmental effects, using multiple gravitational-wave events.

This article is organized as follows. In Sect.~\ref{sec:2}, we will calculate the \ac{ppE} frequency domain waveform correction of a smaller black hole inspiraling a \ac{KBR} black hole. We also consider another black hole with external magnetic field, the \ac{KBM} black hole parallelly and obtain the waveform correction which modifies existing non-spinning result. In Sect.~\ref{sec:3} we perform Fisher parameter estimation of binary inspiral with a magnetic black hole. Moreover, in Sect.~\ref{sec:4} possible degeneracy between the  magnetic corrections and environmental effect are discussed, \textcolor{black}{where we give the relation of parameters between them and implement the statistic $F$ and method to distinguish external magnetic field and environments}. Finally we draw a conclusion in Sect.~\ref{concl}. We adopt $G=c=1$ convention throughout this work.

\section{The waveform correction of binary inspiral with a magnetic blackhole}\label{sec:2}

\subsection{Binary black hole with a \ac{KBR} black hole}\label{kbr}
Magnetic field may influence the orbital evolution of binary black holes and the \ac{GW}s emitted, and the newly found \ac{KBR} spacetime can provide an external magnetic field, intrinsically from spacetime with metric \cite{Podolsky:2025tle}
\begin{eqnarray}
	\mathrm{d}s^2 &=& \frac{1}{\Omega^2} \left[ -\frac{Q}{\rho^2} \left( \mathrm{d}t - a \sin^2\theta \mathrm{d}\phi \right)^2 + \frac{\rho^2}{Q} \mathrm{d}r^2 + \frac{\rho^2}{P} \mathrm{d}\theta^2 \right.\nonumber\\
	&+& \left.\frac{P}{\rho^2} \sin^2\theta \left( a \mathrm{d}t - (r^2 + a^2) \mathrm{d}\phi \right)^2 \right]
\end{eqnarray}
where
\begin{eqnarray}
	\rho^2 &=& r^2 + a^2\cos^2\theta,\\
	P &=& 1 + B^2\left( m^2\frac{I_2}{I_1^2} - a^2 \right)\cos^2\theta,\\
	Q &=& (1 + B^2 r^2)\Delta,\\
	\Omega^2 &=& (1 + B^2 r^2) - B^2\Delta\cos^2\theta,\\
	\Delta &=& \left( 1 - B^2 m^2 \frac{I_2}{I_1^2} \right)r^2 - 2m \frac{I_2}{I_1} r + a^2,\\
	I_1 &=& 1 - \frac{1}{2}B^2 a^2, \quad I_2 = 1 - B^2 a^2.
\end{eqnarray}
with $m,a,B$ denoting the mass, spin and strength of its external magnetic field respectively. Here we consider a smaller blackhole $m_2$ inspiraling into a larger one $m_1$, which is a \ac{KBR} blackhole. For simplicity, we restrict our analysis to the quasi-circular motion of $m_2$ in the equatorial plane of $m_1$, so the only non-vanishing metric components are
\begin{eqnarray}
	g_{tt} &=& -\frac{1}{\Lambda} \left( \frac{\Lambda\Delta}{r^2} - \frac{a^2}{r^2} \right), \\
	g_{t\phi} &=& \frac{1}{\Lambda} \left( \frac{a\Lambda\Delta}{r^2} - \frac{a(r^2 + a^2)}{r^2} \right), \\
	g_{\phi\phi} &=& \frac{1}{\Lambda} \left( \frac{(r^2 + a^2)^2}{r^2} - \frac{a^2\Lambda\Delta}{r^2} \right),
\end{eqnarray}
with $\Lambda(r)\equiv1+B^2r^2$. The orbital angular frequency $\Omega_\phi={\mathrm{d}\phi\over\mathrm{d} t}$ can be determined by extremum condition of the effective potential, in the case of generic stationary and axisymmetric spacetime, we have \cite{Chandrasekhar:1984siy},
\begin{equation}
	\Omega_\phi(r; B) = \frac{-g_{t\phi}' + \sqrt{(g_{t\phi}')^2 - g_{tt}' g_{\phi\phi}'}}{g_{\phi\phi}'},\label{modkpl}
\end{equation}
where $\prime$ represents taking derivative with respect to $r$. Normalization condition $u^\mu u_\mu=-1$ gives the specific energy
\begin{equation}
	\mathcal{E} = -\frac{g_{tt} + \Omega_\phi g_{t\phi}}{\mathcal{N}},
\end{equation}
where the normalization factor is
\begin{equation}
	\mathcal{N} = \sqrt{-\left( g_{tt} + 2\Omega_\phi g_{t\phi} + \Omega_\phi^2 g_{\phi\phi} \right)}
\end{equation}
\subsection{Frequency domain waveform of the binary influenced by \ac{KBR} magnetic field}
The system is equivalent to a reduced mass $\mu\equiv {m_1m_2\over m_1+m_2}$ inspiraling the total mass $M=m_1+m_2$, and $\mu=\eta M$ with $\eta\equiv{m_1m_2\over M^2}$ being the symmetric mass ratio. Then the evolution of orbital frequency $f$ can be computed via
\begin{equation}
	{\mathrm{d}f\over\mathrm{d}t}={1\over\pi}{\mathrm{d}\Omega_\phi\over\mrd E}{\mrd E_\tx{GW}\over\mrd t}\label{dft}
\end{equation}
where
\begin{eqnarray}
	{\mathrm{d}\Omega_\phi\over\mrd E}={\mrd\Omega_\phi\over\mrd r}{\mrd r\over \mu\mrd \mathcal{E}}={\mrd\Omega_\phi\over\mrd r}{\mrd r\over \eta M\mrd \mathcal{E}}\label{dom}
\end{eqnarray}
and we approximate the flux of gravitational radiation with the quadrupole formula
\begin{equation}
	{\mrd E_\tx{GW}\over\mrd t}=-{32\over 5}\eta^2(M\pi f)^{10\over 3}.\label{degw}
\end{equation}
Note \eqref{dom} should use the modified Kepler's law to be expressed as function of $f$, which give us as
\begin{equation}
	r(f)\approx\left(\frac{M}{\pi ^2 f^2}\right)^{1\over 3} \left(1+\frac{2B^2 M^{2/3}}{3 \pi ^{4/3} f^{4/3}}\right).
\end{equation}
Substituting \eqref{degw} into \eqref{dft} to obtain $\dot{f}$, then the \ac{GW} phase taking into account \ac{KBR} spacetime correction can be computed under the stationary phase approximation as
\begin{eqnarray}
	\Psi(f)&\approx& 2\pi ft-2\int\pi f\mrd t\\
	&=&2\pi f\int \left({\mrd f\over \mrd t}\right)^{-1}\mrd f-2\pi\int\left({\mrd f\over \mrd t}\right)^{-1}f\mrd f\nonumber\\
	& &\label{dftjf}\\
	&=&\Psi_\text{vac}(f)+\delta\Psi_\text{KBR}(f),
\end{eqnarray}
where $\Psi_\tx{vac}={3\over 128(\pi Mf)^{{5\over 3}}\eta}$ and
\begin{eqnarray}
	\delta\Psi_\text{KBR}(f)\approx \frac{5 B^2 \left(110 M^{2/3}-81  a (\pi f)^{1\over 3}\right)}{38016 \pi ^3 f^3 \eta  M^{5/3}}.\label{dpskbr}
\end{eqnarray}
When computing $\dot{f}^{-1}$ prior to integrating over $f$ in \eqref{dftjf}, we have retained the leading-order terms in $B$ and $a$ in the small-frequency limit. Note that the first term inside the brackets in the \ac{KBR} correction \eqref{dpskbr} is of $-2$ PN order relative to $\Psi_\text{vac}$, whereas the second term corresponds to a $-1.5$ PN contribution induced by the spin of the \ac{KBR} black hole.

\subsection{Waveform correction of binary system with a \ac{KBM} black hole}

Another black hole solution with external magnetic field is the \ac{KBM} black hole that has line element \cite{Ernst:1976bsr}
\begin{eqnarray}
	\mrd s^2 &=& |\Lambda|^2 \Sigma \left[ -\frac{\Delta}{\mathcal{A}} \mrd t^2 + \frac{\mrd r^2}{\Delta} + \mrd \theta^2 \right] \nonumber\\
	& +& \frac{\mathcal{A}}{\Sigma |\Lambda|^2} \sin^2\theta (\mrd \phi - \omega \mrd t)^2,
\end{eqnarray}
where
\begin{eqnarray}
	\Sigma &=& r^2 + a^2 \cos^2\theta,  \mathcal{A} = (r^2 + a^2)^2 - \Delta a^2 \sin^2\theta, \\
	\Delta &=& r^2 - 2Mr + a^2,  \Lambda = 1 + \frac{1}{4} B^2 \frac{\mathcal{A}}{\Sigma} \sin^2\theta \nonumber\\
	& -& \frac{i}{2} B^2 M a \cos\theta \left( 3 - \cos^2\theta + \frac{a^2}{\Sigma} \sin^4\theta \right),
\end{eqnarray}
and 
\begin{widetext}
\begin{eqnarray}
	\omega = \frac{a}{r^2 + a^2} \Bigg\{ &\left(1 - B^4 M^2 a^2\right) - \Delta \Bigg[ \frac{\Sigma}{\mathcal{A}} + \frac{B^4}{16} \bigg( -8Mr \cos^2\theta (3 - \cos^2\theta) - 6Mr \sin^4\theta \nonumber\\
	&+ \frac{2Ma^2 \sin^6\theta}{\mathcal{A}} \left[ r(r^2 + a^2) + 2Ma^2 \right] + \frac{4M^2 a^2 \cos^2\theta}{\mathcal{A}} \left[ (r^2 + a^2)(3 - \cos^2\theta)^2 - 4a^2 \sin^2\theta \right] \bigg) \Bigg] \Bigg\}.
\end{eqnarray}
\end{widetext}
and another version of it was given in \cite{Aliev:1989wz}. Like in the \ac{KBR} case above, we consider a smaller black hole inspiraling to one \ac{KBM} black hole in the equatorial plane and calculate the modified Kepler law \eqref{modkpl}, the evolution of frequency \eqref{dft} and the waveform correction \textcolor{black}{which} turns out to be 
\begin{eqnarray}
	\delta\Psi_\text{KBM}(f)&\approx& \frac{5 B^2 \left(783 \pi  a f+55 \pi ^{2/3} (f M)^{2/3}+216\right)}{76032 \pi ^{11/3} f^{11/3} \eta  M^{5/3}}.\nonumber\\
	& &\label{dpskbm}
\end{eqnarray}
Its leading-order term is at $-3$ PN, and is approximately three times the simpler nonspinning result reported in the earlier work \cite{Barausse:2014tra}. The spinning contribution is $1.5$ PN order relative to the leading correction, with a 1 PN term lying between them. The result is the same as that of another version \cite{Aliev:1989wz} of KBM black hole because of common modified Kepler law with \cite{Iyer:2025uyb} up to order $B^2$ and $a$.

\subsection{Considering \ac{ppE} correction with higher modes and sub-leading terms}


If the binary is highly asymmetric, higher modes becomes relevant, so the \ac{ppE} waveform reads \cite{gjmfk},
\begin{equation}
	\td{h}(f)=\sum_{lm}\td{h}_{lm}(f)=\sum\td{h}^\text{GR}_{lm}(f)e^{i\delta\Psi_{lm}}.\label{hlm}
\end{equation}
In this article, IMRPhenomXHM \textcolor{black}{\cite{phexhm}} is used to obtain the \ac{GR} waveform with higher modes.
Because the amplitude correction is nearly negligible, we will only consider the phase correction,
\begin{equation}
	\delta\Psi_{lm}=\beta_{lm}u^{b_{lm}}.\label{dpslm}
\end{equation}
where $u=(\pi\mc{M}_cf)^{1/3}$, and $\mc{M}_c=(m_1m_2)^{3/5}/(m_1+m_2)^{1/5}$ is the chirp mass for the binary with component masses $m_1$ and $m_2$.
The corrections for different higher modes can be represented by that of $22$ mode as \cite{gjmfk}
\begin{equation}
	\beta_{lm}=\left(\frac{2}{m}\right)^{\frac{b_{lm}}{3}-1}\beta_{22},~~~b_{lm}=b_{22}.\label{betalm}
\end{equation}

When sub-leading terms of correction are taken into account, \eqref{dpslm} should be modified accordingly as
\begin{equation}
	\delta\Psi_{lm}=\beta_{lm}u^{b_{lm}}+\beta_{1lm}u^{b_{1lm}}+\cdots.\label{dpslmj}
\end{equation}

\section{Parameter estimation}\label{sec:3}

In this section we will consider parameter estimation of the magnetic strength $B$ that influences \ac{GW} from massive binary black holes with Tianqin detector.
\subsection{Fisher information matrix}

The sky-averaged noise \ac{PSD} for TianQin \textcolor{black}{\cite{Wang:2019ryf}} is 
\begin{eqnarray}
	S_n(f)&=&\frac{10}{3}\frac{1}{L^2}\left[1+\left(\frac{2fL_0}{0.41c}\right)^2\right]\nonumber\\
	&\times&\left[\frac{4S_a}{(2\pi f)^4}\left(1+\frac{10^{-4}\text{Hz}}{ f}\right)+S_x\right]
\end{eqnarray}
with the \textcolor{black}{arm length} of TianQin $L=\sqrt{3}\scf{5}\text{km}$, $\sqrt{S_a}=10^{-15}\text{m}\cdot\text{s}^{-2}\text{Hz}^{-1/2}$ and $\sqrt{S_x}=10^{-12}\text{m}\cdot\text{Hz}^{-1/2}$ are the acceleration and position noise of TianQin, respectively.

Next, with waveform \eqref{hlm} given by \eqref{dpskbr} or \eqref{dpskbm}, the \textcolor{black}{\ac{PE} precision} of the respective parameter $\theta^i$ can be given by $\delta\theta^i=\sqrt{\Sigma_{ii}}$ and 
\begin{equation}
	\Gamma_{ij}=2\int_{f_\text{min}}^{f_\text{max}}\frac{\partial_ih(f)\partial_jh^\ast(f)+\partial_ih^\ast(f)\partial_jh(f)}{S_n(f)}df
\end{equation}
 in the large \ac{SNR} limit, with the partial derivative $\partial_i$ corresponding to the $i$-th parameter of $\hat\theta=(\mc{M}_c,\eta,\chi_1,\chi_2,B,D_L)$.
Polarization and inclination angle are not taken into account and we have averaged the orientation of the sources.
Since the \ac{ppE} correction is used for the inspiral stage, we will only use the inspiral signal to do the analysis, and thus we will have
\begin{equation}
	f_\text{max}=f_\text{ISCO}=\frac{1}{6\sqrt{6}\pi M}
\end{equation}
and
\begin{equation}
	f_\text{min}=\left(\frac{5}{256}\right)^{3/8}\frac{1}{\pi}\mc{M}_c^{-5/8}T^{-3/8}
\end{equation}


\subsection{The \textcolor{black}{\ac{PE} precision} of magnetic strength $B$}
For the parameters of the sources, we choose $\chi_1=0.9,\chi_2=0.8,T=5\text{yr},\phi_c={\pi\over 4},\psi=0,\iota={\pi\over 4}$.
$\mc{M}_c$ varies between $10^4\sim10^7M_\odot$, and $\eta$ varies between $0\sim 0.25$

\begin{figure}
	\begin{center}
		\includegraphics[scale=0.55]{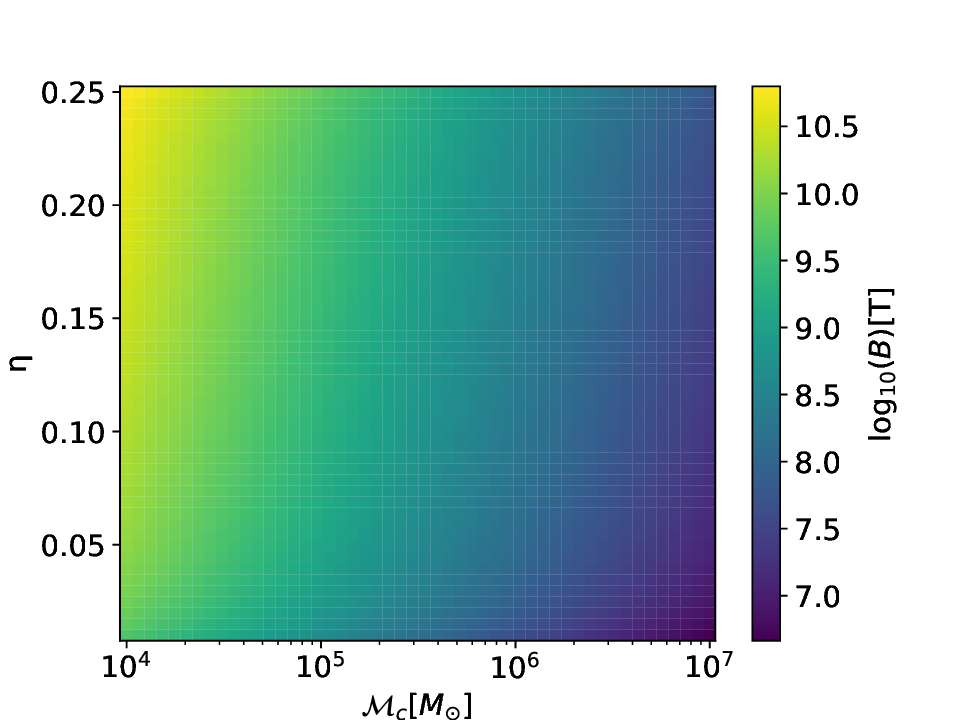}
		\caption{The magnetic strength $B=0.1B_\tx{extr}$ of different systems.}\label{derhemkbrB}
	\end{center}
\end{figure}

\begin{figure}
	\begin{center}
		\includegraphics[scale=0.55]{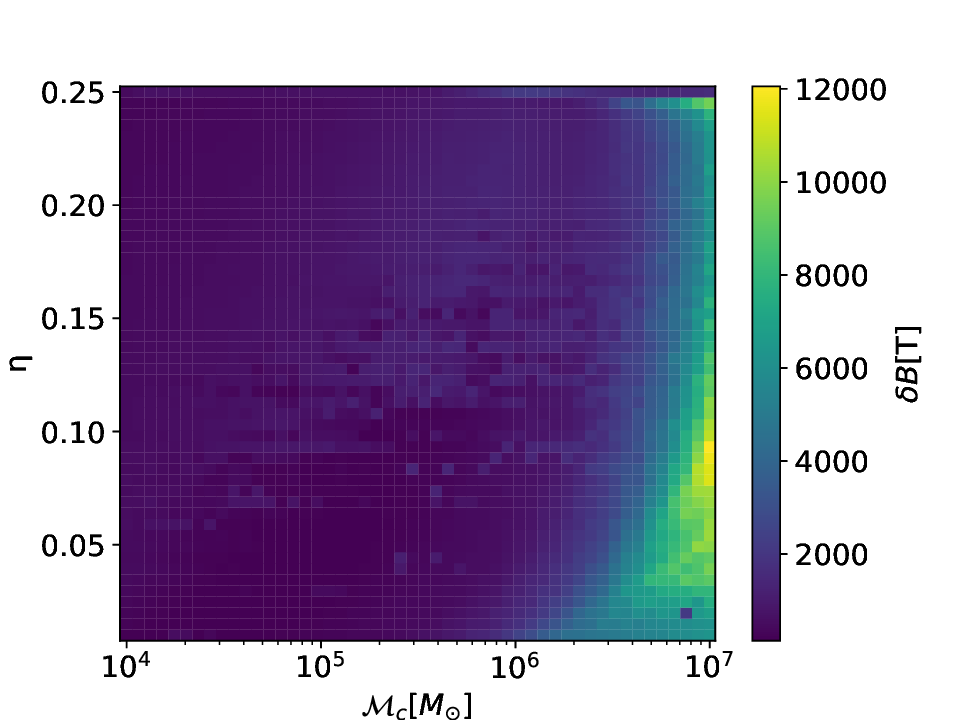}
		\caption{The precision to measure magnetic strength $\delta B$ of different systems.}\label{derhemkbr}
	\end{center}
\end{figure}

From the \ac{KBR} waveform correction \eqref{dpskbr}, the \textcolor{black}{\ac{PE} precision} of $B$ can be calculated by the Fisher matrix, and we consider Tianqin sensitivity curve. \figref{derhemkbrB} and \figref{derhemkbr} show the magnetic strength $B$ and the \textcolor{black}{\ac{PE} precision} $\delta B$ of different systems located at $D_L=1\text{Gpc}$, where we have adopted the dimensionless spin parameter of the larger black hole $\chi=0.9$, and $B$ is 0.1 times the extremal value $B_\text{ex}$ given by spin $\chi$ as \cite{Podolsky:2025tle}
\begin{eqnarray}
	B_\text{ex,BR}&=&\left(2\chi^{-4}m_1^{-2}(1-\sqrt{1-\chi^2})\sqrt{1-\chi^2}\right)^{1\over 2},\label{bex}\\
	\chi&\equiv&{J\over m^2}={a\over m}={2a\over R_s}.
\end{eqnarray}
For $m_1=10^6M_\odot$ and $a=0.9$, $B_\text{ex,BR}=7.23\times 10^9\text{T}$. 
\begin{figure}
	\begin{center}
		\includegraphics[scale=0.55]{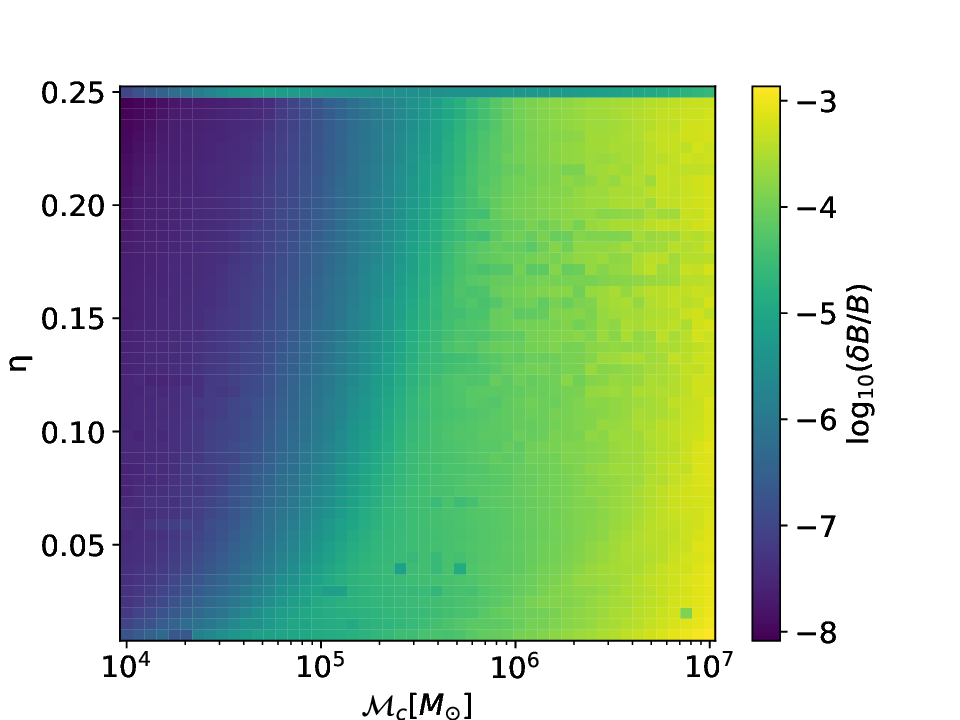}
		\caption{Relative precision of magnetic strength $\delta B/B$ of different systems.}\label{derhemkbrxd}
	\end{center}
\end{figure}
From \figref{derhemkbr}, it is shown smaller $\mc{M}_c$ and larger $\eta$ system has smaller $\delta B$, but $B$ is larger as seen from \figref{derhemkbrB}.  Therefore, in \figref{derhemkbrxd} the relative precision becomes better for smaller $\mc{M}_c$ and larger $\eta$ system with $\delta B/B$ down to order of $10^{-8}$.

\begin{figure}
	\begin{center}
		\includegraphics[scale=0.55]{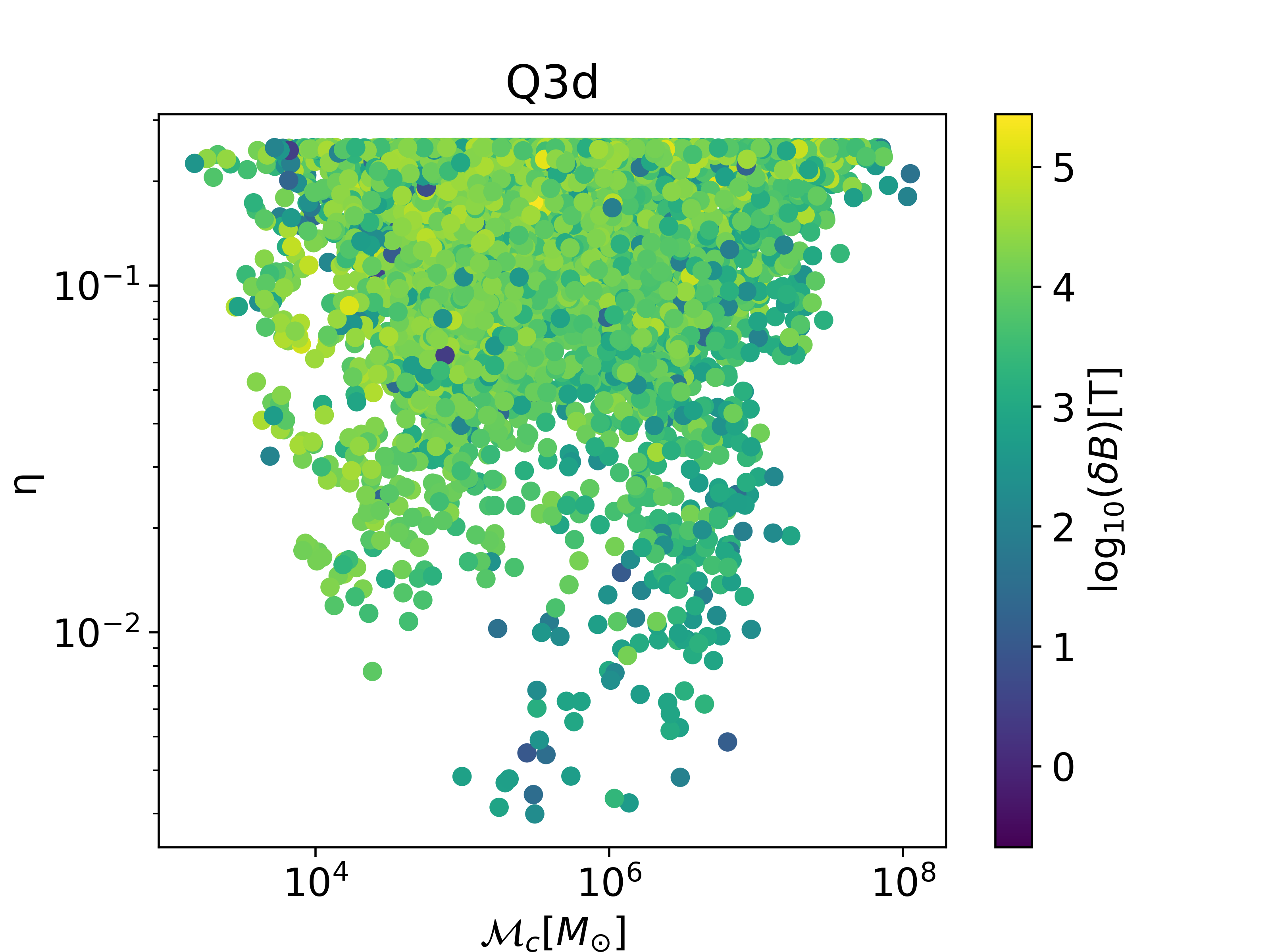}
		\caption{The precision to measure magnetic strength $\delta B$ of different systems, considering Q3d sources.}\label{emq3dkbr}
	\end{center}
\end{figure}

\begin{figure}
	\begin{center}
		\includegraphics[scale=0.55]{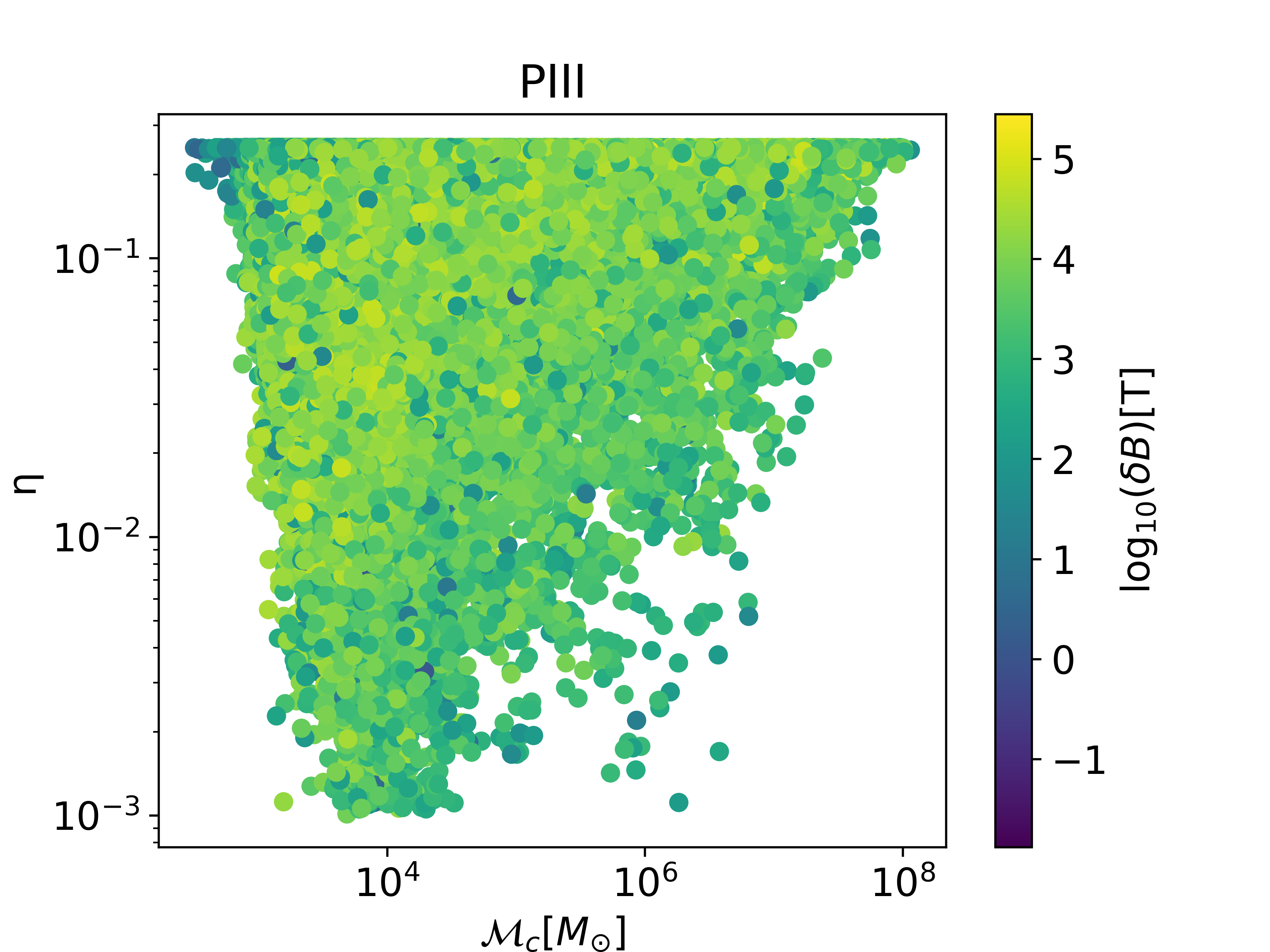}
		\caption{The precision to measure magnetic strength $\delta B$ of different systems, considering PIII sources.}\label{emp3kbr}
	\end{center}
\end{figure}

\begin{figure}
	\begin{center}
		\includegraphics[scale=0.55]{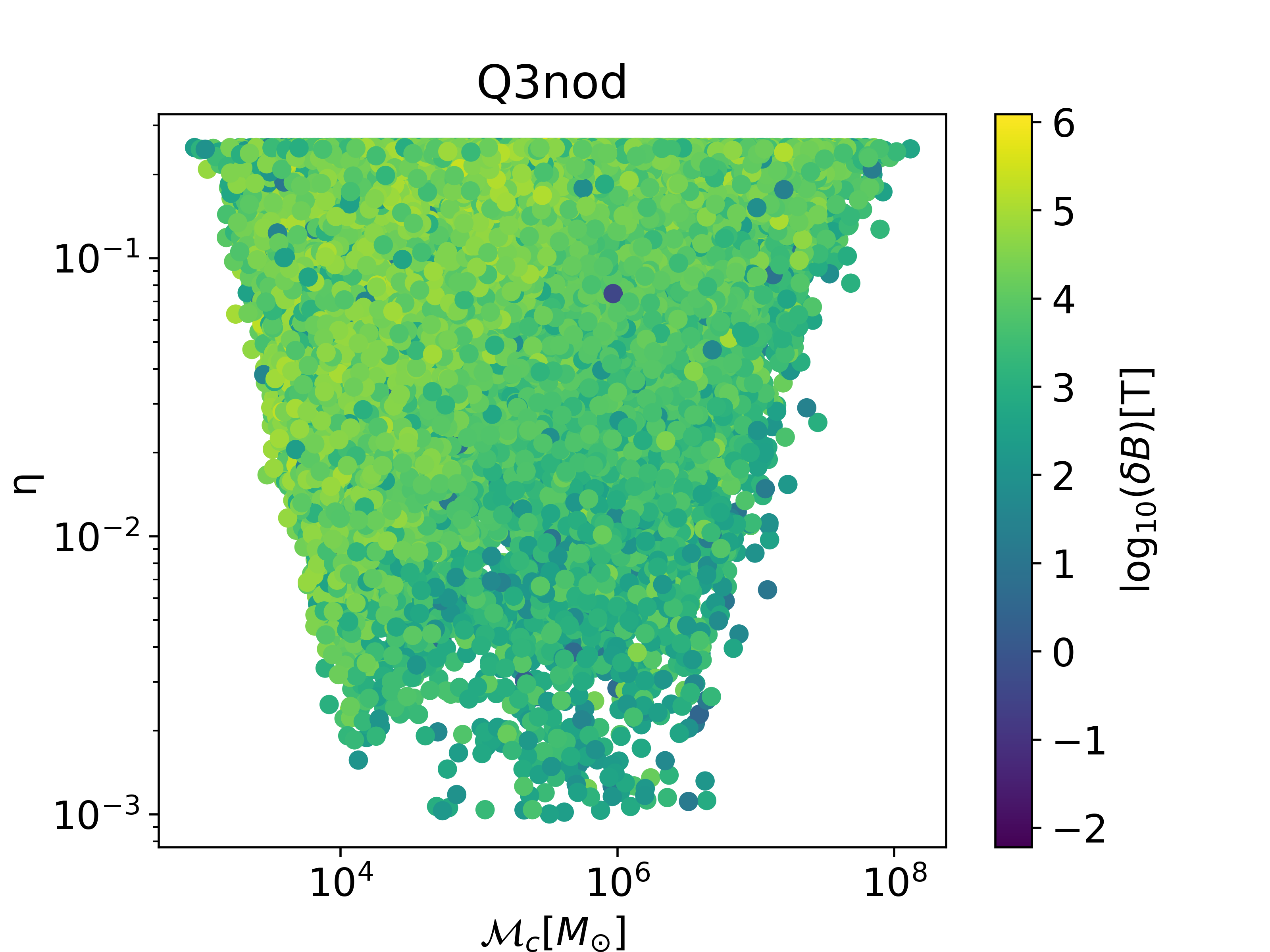}
		\caption{The precision to measure magnetic strength $\delta B$ of different systems, considering Q3nod sources.}\label{emq3nodkbr}
	\end{center}
\end{figure}

We also calculate the result considering the three models for massive black hole binaries \cite{Barausse:2012fy,Sesana:2014bea,Antonini:2015sza,Klein:2015hvg},
two heavy-seed models Q3d and Q3nod, and a light-seed model popIII. Each model has 1000 mock catalogs of sources containing 18112, 271444, and 56618 events detectable for TianQin \cite{Wang:2019ryf} five-year observation with parameters $(z,m_1,m_2,\chi_1,\chi_2,\iota)$.
The measuring precision is plotted in \figref{emq3dkbr}, \figref{emp3kbr} and \figref{emq3nodkbr}. Generally speaking, PIII has the best constraints of magnetic strength $B$ at 0.1 the extreme $B$ value, and Q3d the worst. Since $\delta B$ is smaller for lower mass system, the light seed model PIII has the best result(medium).
While Q3nod has the largest number of sources, its $\delta B$ can be as down to $10^{-2}\text{T}$ but the upper bound is also higher. 

\begin{figure}
	\begin{center}
		\includegraphics[scale=0.55]{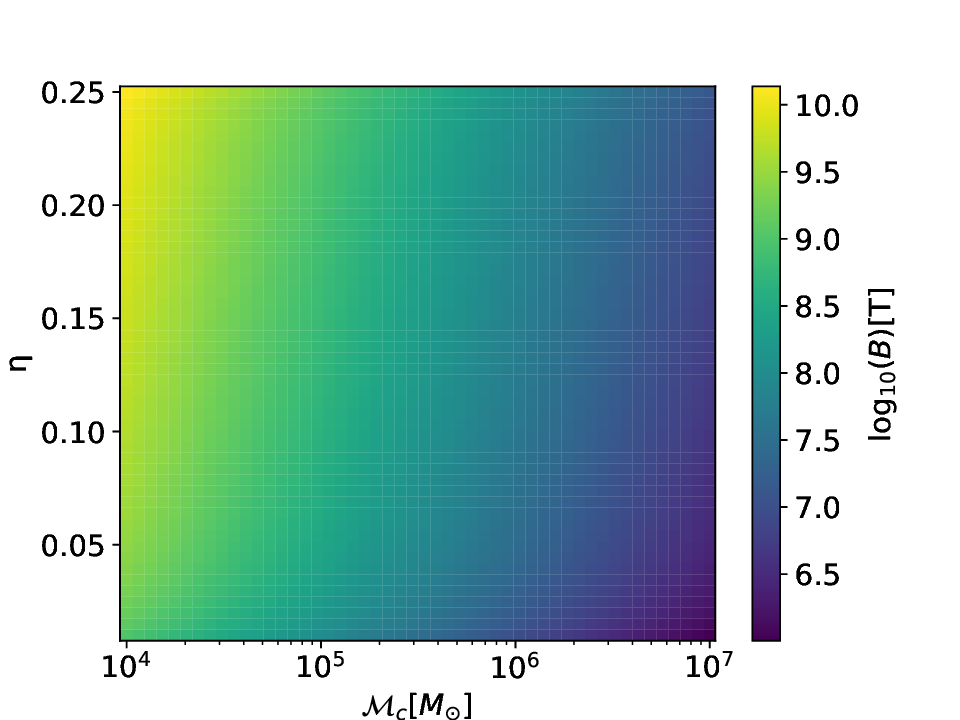}
		\caption{The magnetic strength $B=0.1B_\tx{extr}$ of different systems with a \ac{KBM} blackhole.}\label{derhemkbmB}
	\end{center}
\end{figure}
\begin{figure}
	\begin{center}
		\includegraphics[scale=0.55]{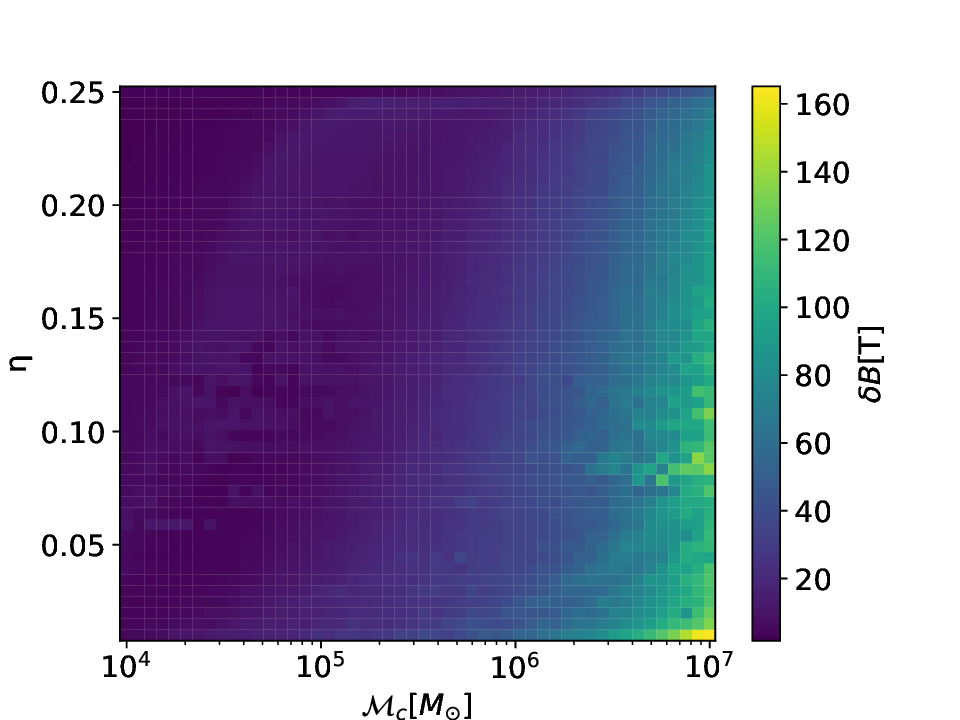}
		\caption{The precision to measure magnetic strength $\delta B$ of different systems with a \ac{KBM} blackhole.}\label{derhemkbm}
	\end{center}
\end{figure}
\begin{figure}
	\begin{center}
		\includegraphics[scale=0.55]{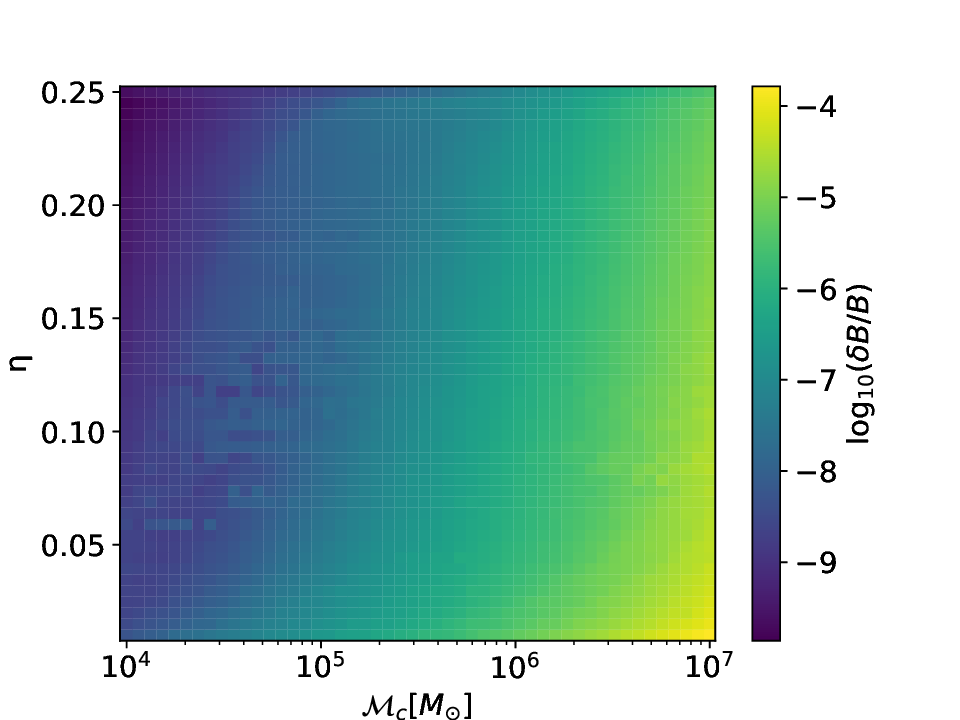}
		\caption{The relative precision to measure magnetic strength $\delta B/B$ of different systems with a \ac{KBM} blackhole.}\label{derhemkbmxd}
	\end{center}
\end{figure}

Then we present the result for KBM black hole. It has an extremal value $B_\text{ex}$ as well, which mildly depends on the black hole's spin but can only be numerically solved \cite{Iyer:2025uyb}. Thus we calculate the measuring precision $\delta B$ at $0.1B_\text{ex}$ in \figref{derhemkbm} but use the spinless $B_\text{ex}$,
\begin{eqnarray}
	B_\text{ex,BM}=0.189m_1^{-1},\label{bexkm}
\end{eqnarray}
plotted in \figref{derhemkbmB}. For $m_1=10^6M_\odot$, $B_\text{ex,BM}=1.58\times 10^9\text{T}$. The $B_\text{ex}$ of BM black hole is about one fifth the value of BR with $a=0.9$ in \eqref{bex}, but BM has better $\delta B$ at $0.1B_\text{ex}$, and relative precision that is plotted in  $\delta B/B$ in \figref{derhemkbmxd}.

\section{Distinguishing spacetime intrinsic magnetic field and environments}\label{sec:4}
\subsection{The relation between gravitational pull and external magnetic fields effect}
As pointed in the introduction, external magnetic effect might be similar to environments. For example, 
the waveform correction due to gravitational pull (GP) of power-law distribution
\begin{eqnarray}
	\rho(r)=\rho_0\left({r_0\over r}\right)^{\gamma}\label{rhlb}
\end{eqnarray}
is \cite{eda,Barausse:2014tra,constr}
\begin{eqnarray}
\delta\Psi_\tx{GP}&=&-{5f^{{2\gamma-11\over 3}}M^{{-\gamma-5\over 3}}\pi^{{2\gamma-11\over 3}}(2\gamma-5)r_0^\gamma\rho_0\over 16(\gamma-3)(2\gamma-11)\eta}\label{dpsgp}
\end{eqnarray}
Comparing it with the leading term of \ac{KBM} \eqref{dpskbm} and \ac{KBR} \eqref{dpskbr}, the power index that causes the same order of corrections of \ac{BM} and \ac{BR} magnetic field are $\gamma_\tx{BM}=0$ and $\gamma_\tx{BR}=1$, respectively. This means the gravitational pull of constant or $\gamma=1$ power-law matter distribution can be degenerate with intrinsic magnetic field of spacetime. \textcolor{black}{It is noteworthy that} the power index $\gamma$ needs not to be steep, which is favored by some theory \cite{ulscdm} or observation \cite{Spekkens:2005ik,Milosavljevic:2001vi} of matter at the galactic centers.

\begin{figure}
	\begin{center}
		\includegraphics[scale=0.55]{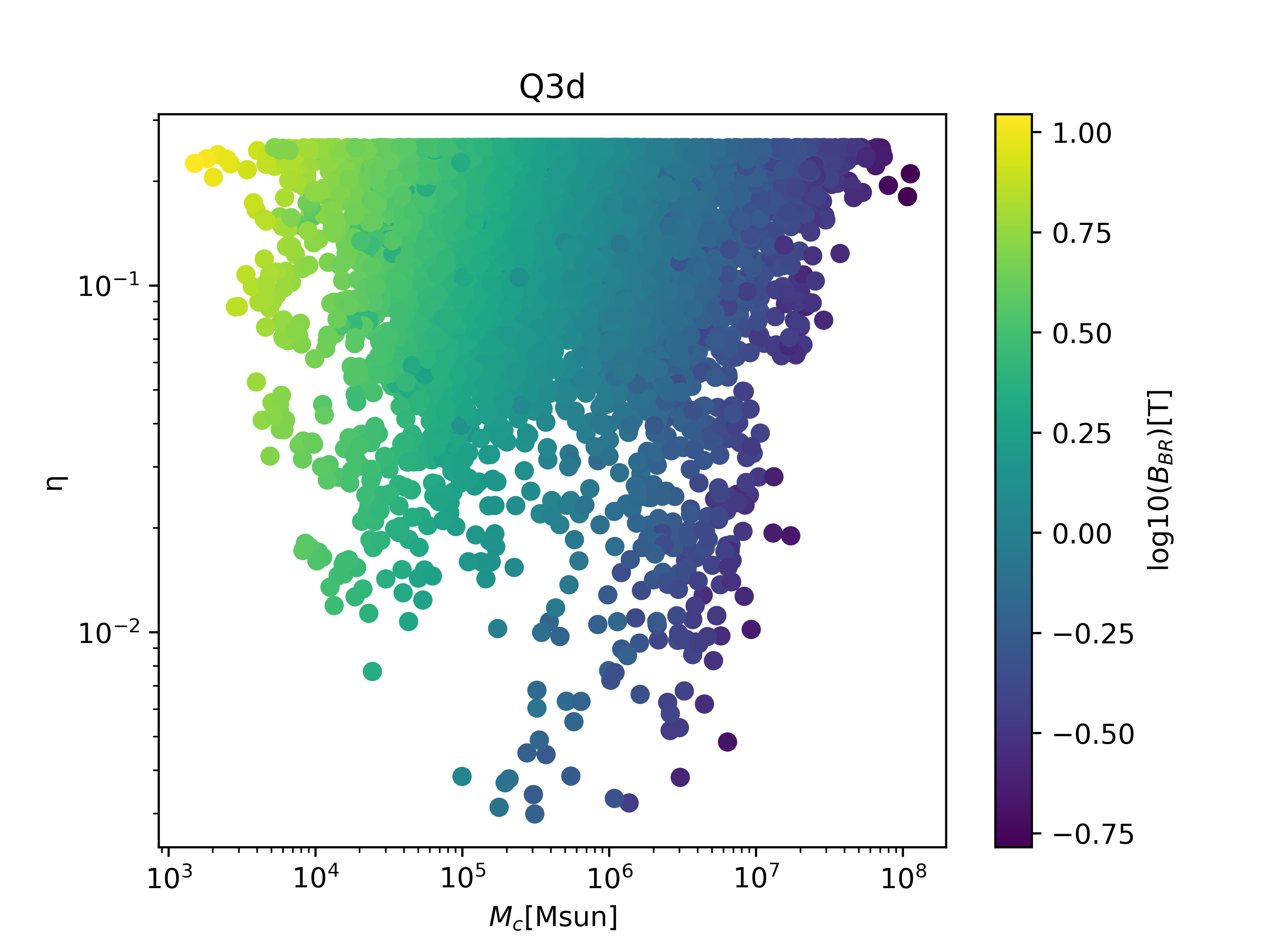}
		\caption{The magnetic strength $B_\tx{BR}$ corresponding to gravitational pull from power-law matter, considering Q3d sources.}\label{emq3drhbkbr}
	\end{center}
\end{figure}

\begin{figure}
	\begin{center}
		\includegraphics[scale=0.55]{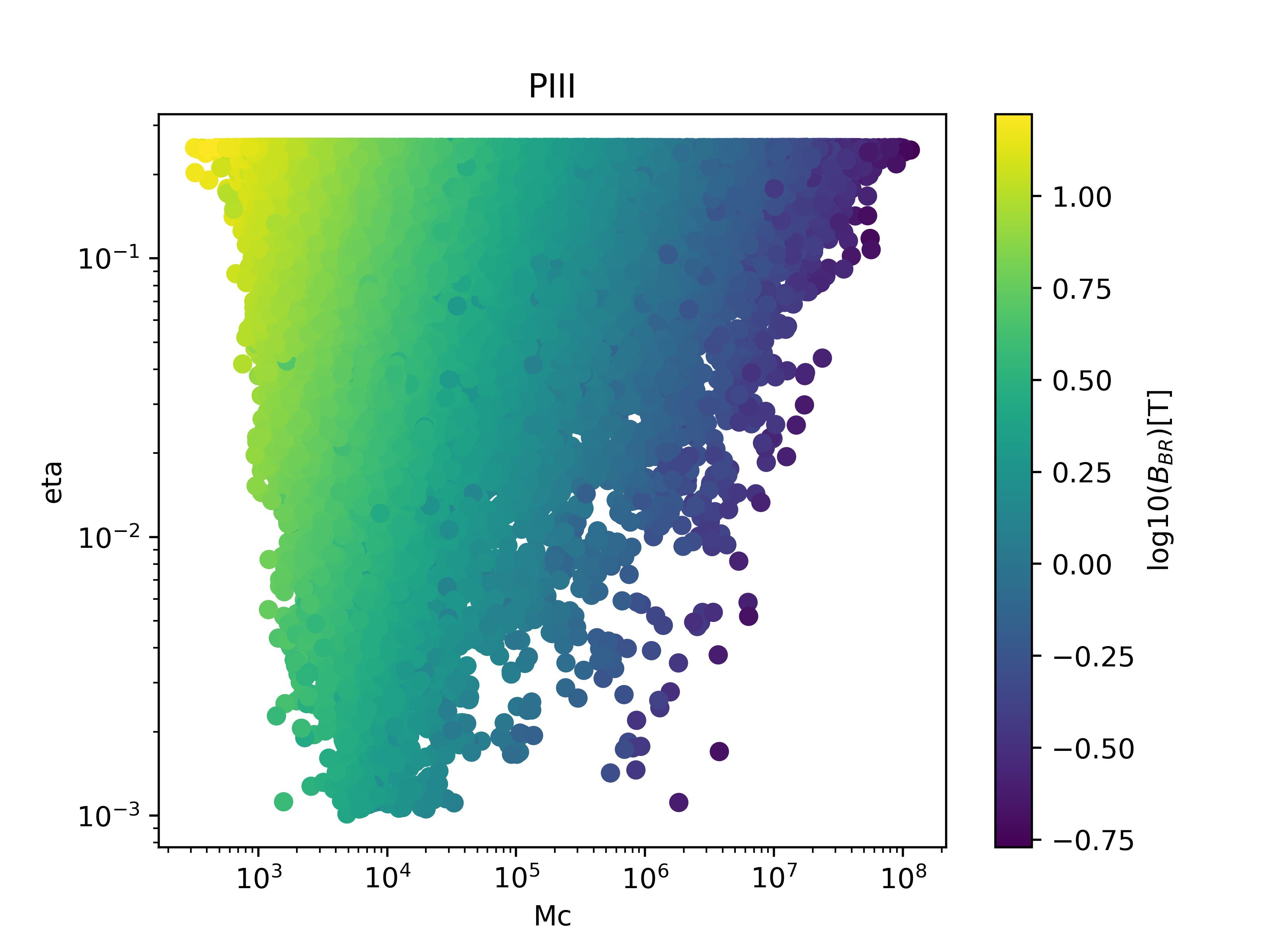}
		\caption{The magnetic strength $B_\tx{BR}$ corresponding to gravitational pull from power-law matter, considering PIII sources.}\label{emp3rhbkbr}
	\end{center}
\end{figure}

\begin{figure}
	\begin{center}
		\includegraphics[scale=0.55]{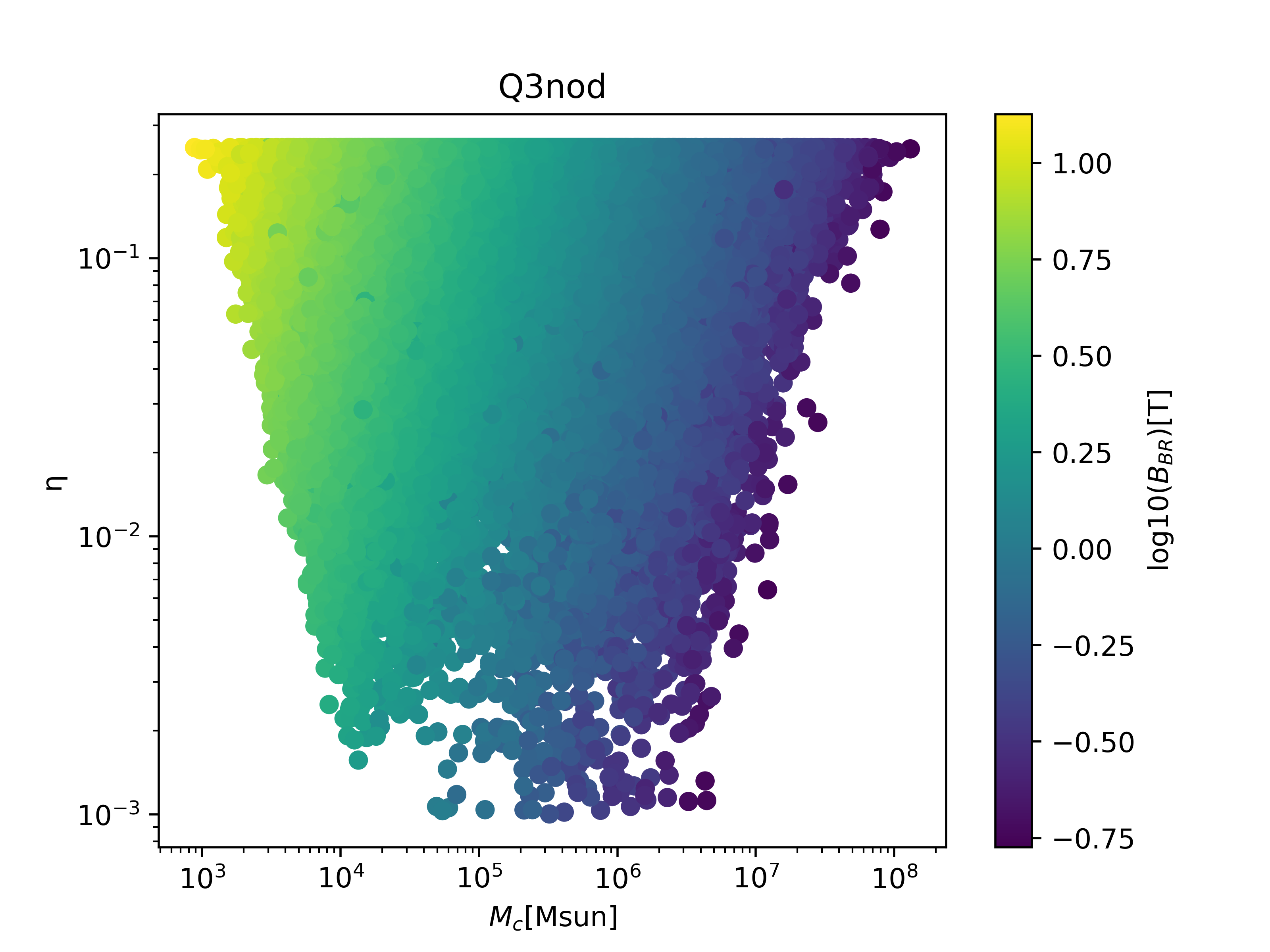}
		\caption{The magnetic strength $B_\tx{BR}$ corresponding to gravitational pull from power-law matter, considering Q3nod sources.}\label{emq3nodrhbkbr}
	\end{center}
\end{figure}

Then it is direct to derive the corresponding relation between density $\rho_0$ and magnetic strength $B$ of \ac{KBR} and \ac{KBM} magnetic fields. Equating \eqref{dpsgp} with $\gamma=1$ to \eqref{dpskbr} yields
\begin{eqnarray}
	B_\tx{BR}=3\sqrt{2 \rho_0r_0\over 5M}.\label{bbr}
\end{eqnarray}
Similarly, setting $\gamma=0$ in \eqref{dpsgp} which has the corresponding magnetic strength,
\begin{eqnarray}
	B_\tx{BM}=\sqrt{10 \rho_0\over 3}.\label{bbm}
\end{eqnarray}
The values of $r_0$ and $\rho_0$ in \eqref{rhlb} are determined by some additional conditions, here we relate them to the 
mass of power-law matter(which is related to the critical density of our universe $\rho_c=9.73\times 10^{-27}\text{kg}/\text{m}^3$  \cite{WMAP1OB} and the \textcolor{black}{redshift} $z$)  and \ac{BH}s' radius of gravitational influence as \cite{yuan2024,pmm,Yuan:2025pbu}:
\begin{eqnarray}
	\rho_0&=&200 \rho_{cm}5^\gamma\left(\frac{N}{2}\right)^\frac{\gamma}{3-\gamma}\label{equ:rho0}\\
	r_0&=&\left(\frac{M(3-\gamma)}{2\pi\rho_0}\right)^\frac{1}{3}5^\frac{\gamma-3}{3}.\label{equ:r0}
\end{eqnarray}
$N$ is generally taken in the range of  $10^3-10^6$ \cite{eda,mdme,galdy,Sesana:2014bea,yuan2024,Yuan:2025pbu}. The combination $\rho_0 r_0^\gamma$ can be simplified:
\begin{equation}
	\rho_0r_0^\gamma=(2\pi)^{-\frac{\gamma}{3}}5^\frac{\gamma(\gamma-3)}{3}
	(3-\gamma)^\frac{\gamma}{3}\mc{M}_c^\frac{\gamma}{3}
	\eta^{-\frac{\gamma}{5}}\rho_0^{1-\frac{\gamma}{3}}.
\end{equation}
We adopt $N=10^6$ in this article. 
 From \eqref{bbr}, the results of $B_\tx{BR}$ for Q3d, PIII and Q3nod sources are plotted in \figref{emq3drhbkbr}, \figref{emp3rhbkbr} and \figref{emq3nodrhbkbr}, so as to obtain the probability distribution of $B_\tx{BR}$ for the three models.
The peaks of $\log_{10}B_\text{BR}$ are $0.23, 0.98$ and $0.35$ for Q3d, PIII, and Q3nod. It is easily seen that PIII has a significantly larger peak value than the other two models, thanks to its relatively smaller \ac{BH} masses.
\begin{figure}
	\begin{center}
		\includegraphics[scale=0.55]{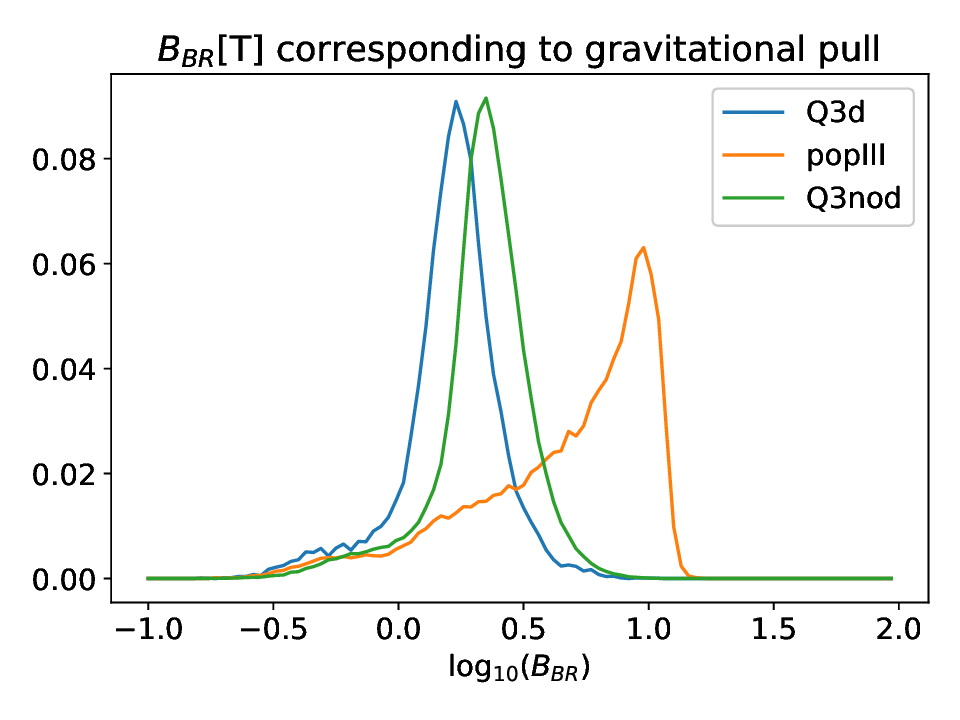}
		\caption{Distribution of $B_\tx{BR}$ for three models.}\label{tedgdrhbkbr}
	\end{center}
\end{figure}

\begin{figure}
	\begin{center}
		\includegraphics[scale=0.55]{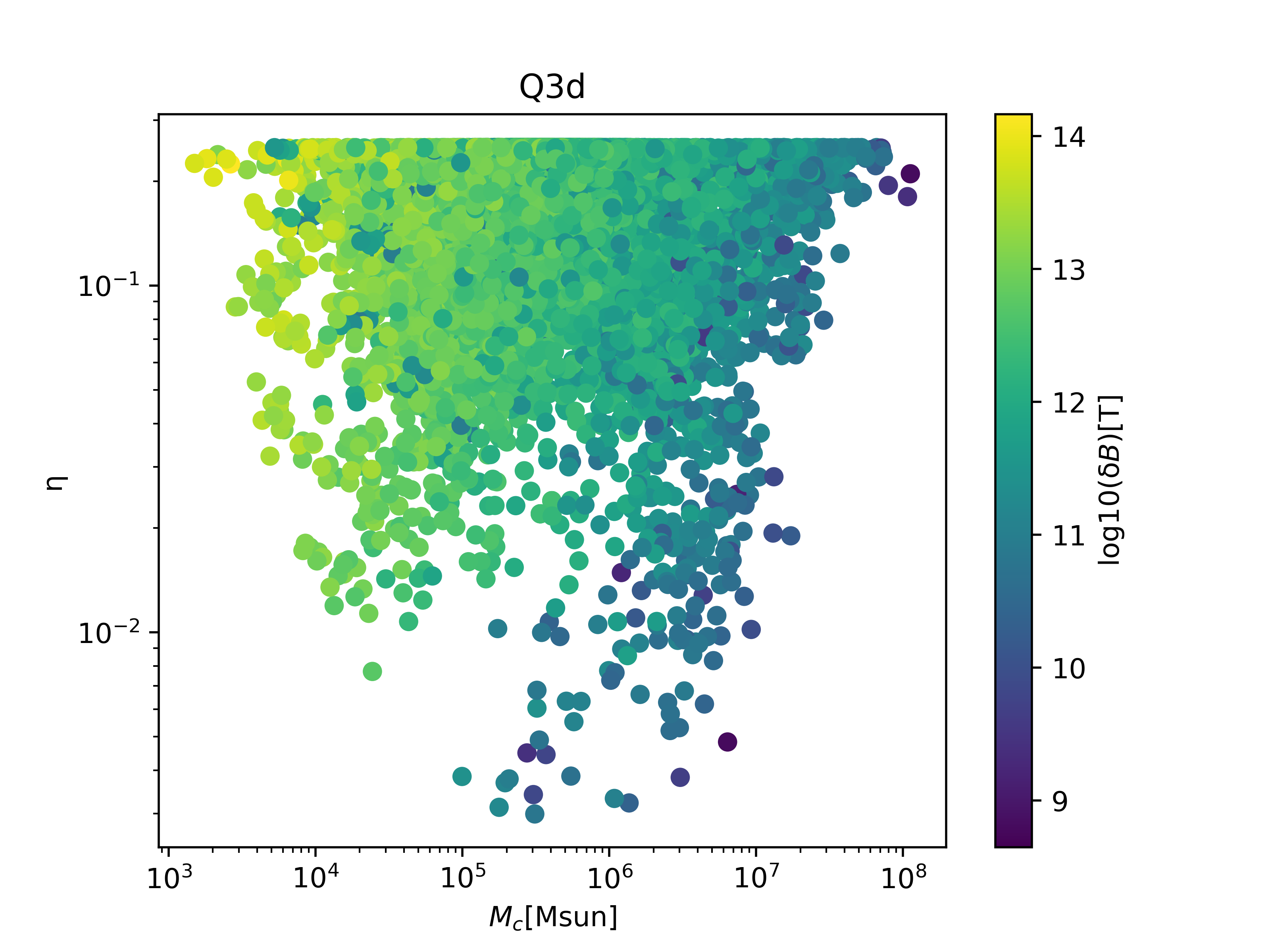}
		\caption{The PE precision $\delta B$ at magnetic strength $B_\tx{BR}$ corresponding to gravitational pull from power-law matter, considering Q3d sources.}\label{emq3drhbkbrdrh}
	\end{center}
\end{figure}
\begin{figure}
	\begin{center}
		\includegraphics[scale=0.55]{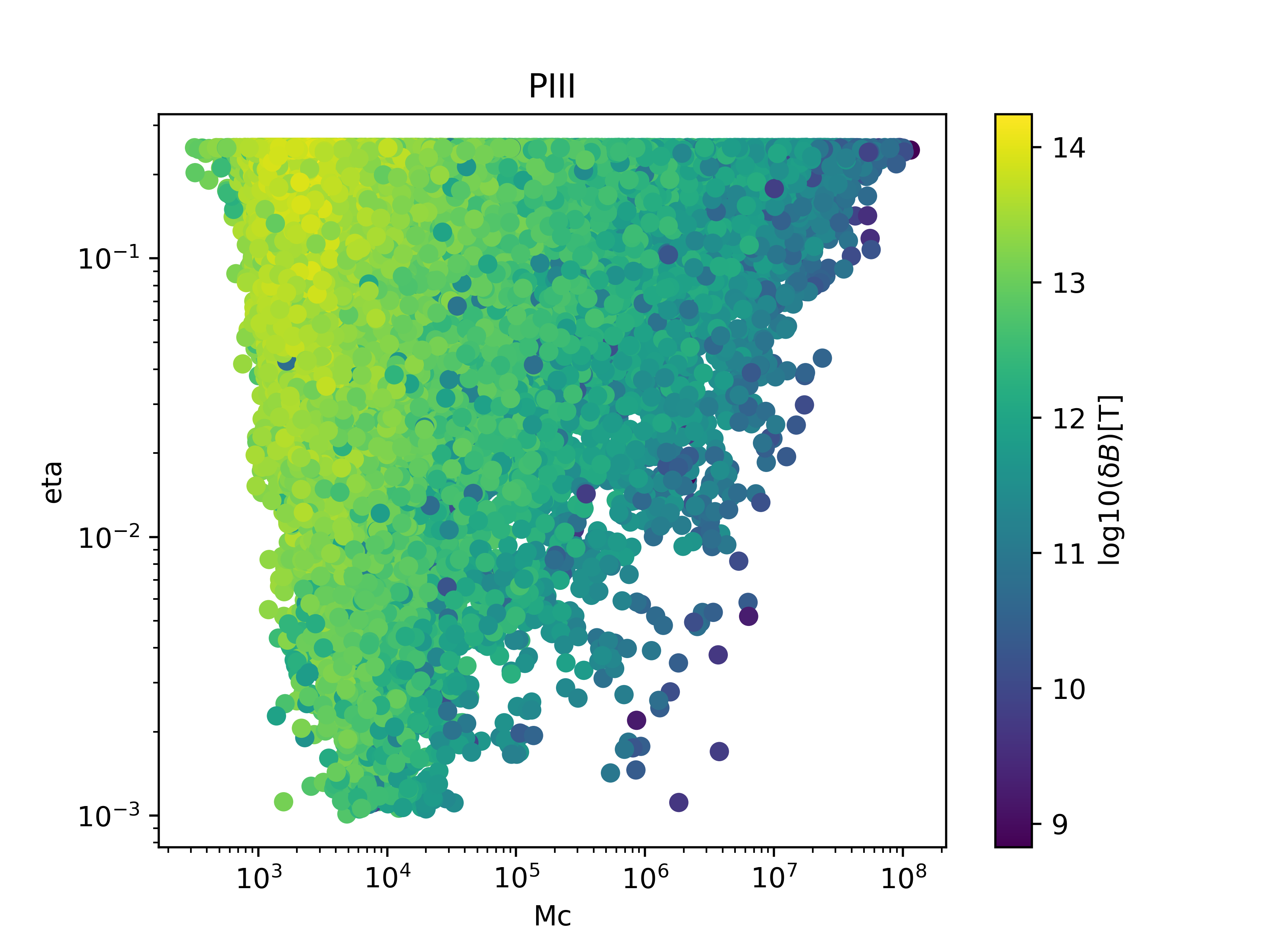}
		\caption{The PE precision $\delta B$ at magnetic strength $B_\tx{BR}$ corresponding to gravitational pull from power-law matter, considering PIII sources.}\label{emp3rhbkbrdrh}
	\end{center}
\end{figure}
\begin{figure}
	\begin{center}
		\includegraphics[scale=0.55]{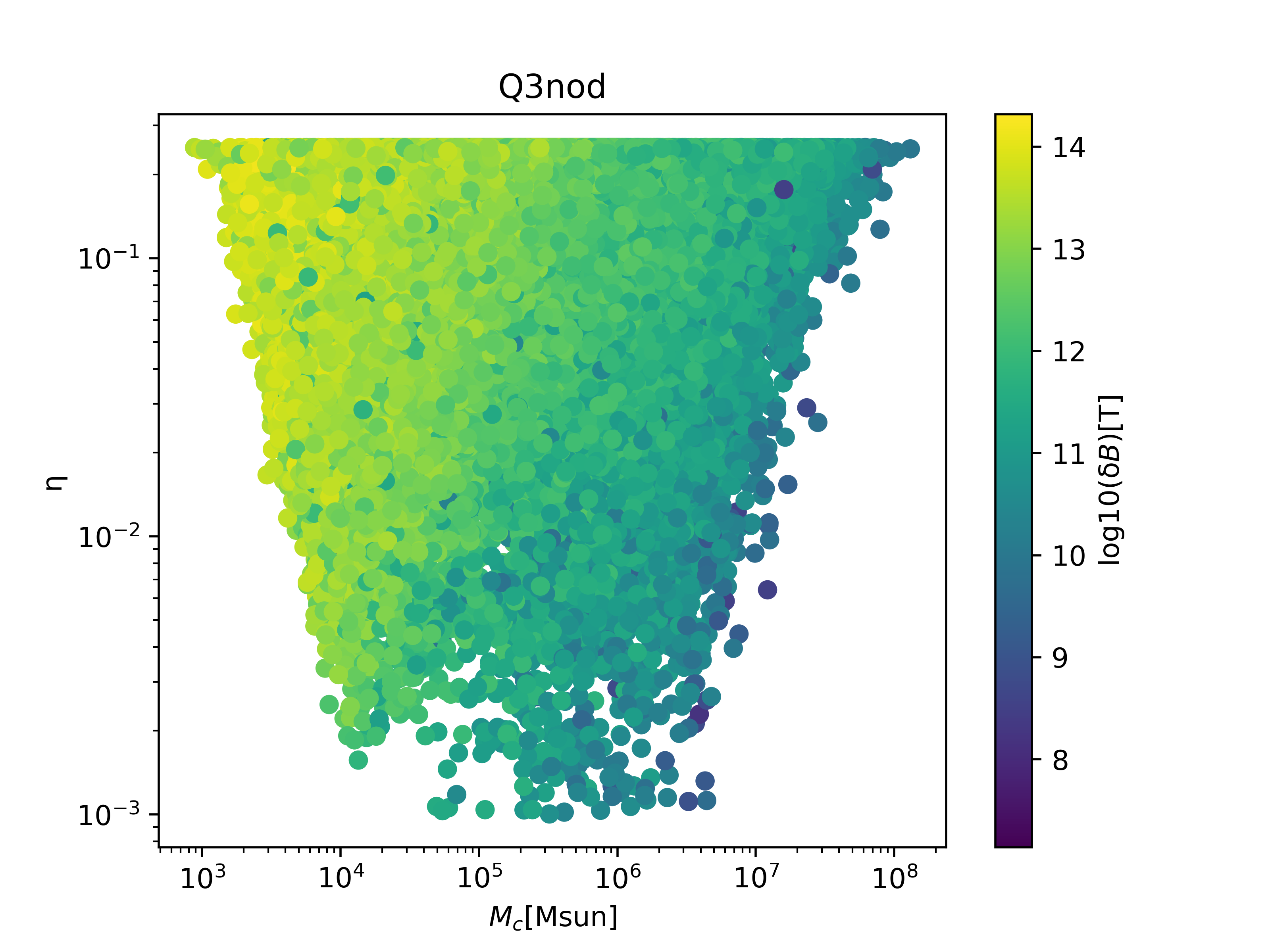}
		\caption{The PE precision $\delta B$ at magnetic strength $B_\tx{BR}$ corresponding to gravitational pull from power-law matter, considering Q3nod sources.}\label{emq3nodrhbkbrdrh}
	\end{center}
\end{figure}
\begin{figure}
	\begin{center}
		\includegraphics[scale=0.55]{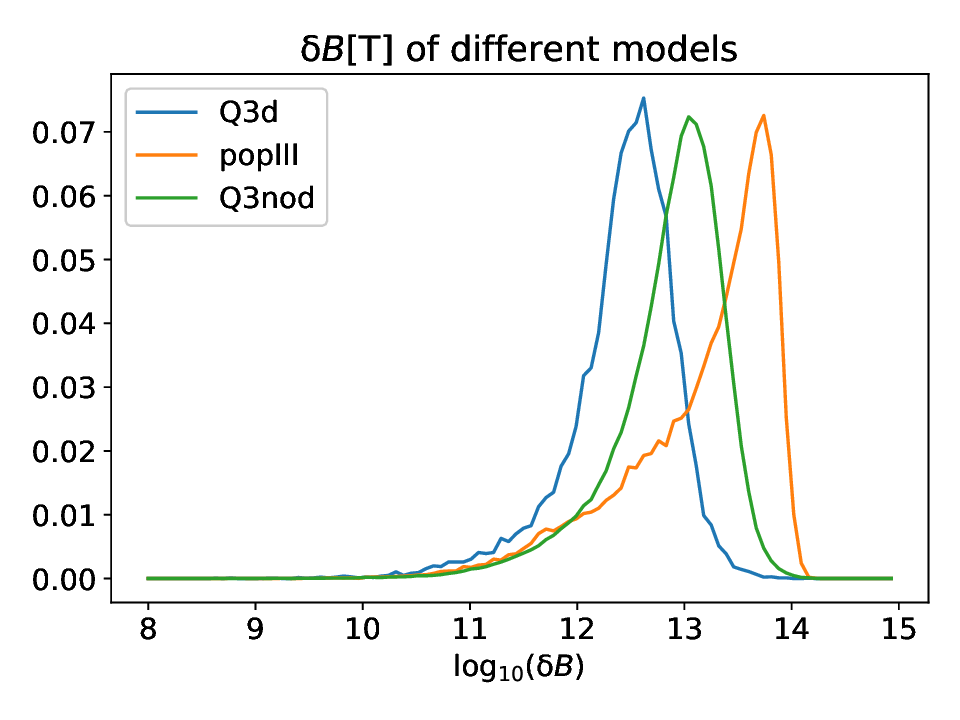}
		\caption{Distribution of $\delta B$ at magnetic strength $B_\tx{BR}$ for three models.}\label{tedgdrhbkbrdrh}
	\end{center}
\end{figure}

Then we calculate the measuring precision $\delta B$ at the magnetic strength corresponding to matter gravitational pull \eqref{bbr}, as shown in \figref{emq3drhbkbrdrh}, \figref{emp3rhbkbrdrh} and \figref{emq3nodrhbkbrdrh} for the three models with probability distribution in \figref{tedgdrhbkbrdrh}. Similar to the distribution of $B_\text{BR}$ in \figref{tedgdrhbkbr}, popIII model has the relatively larger $\delta B$ than the other models, and with larger peak value.

\subsection{Distinguish gravitational pull and Bertotti-Robinson external megnetic field}

In this subsection we intend to distinguish gravitational pull (GP) and real magnetic effects with multiple events. For the real case, we adopt two choices to discuss the distinguishability between it and GP dependent case. The analysis with each choice are then presented in \ref{rdbex} and \ref{smgp}, respectively.

\subsubsection{Random $B$ below the extremal value $B_\text{ex}$}\label{rdbex}
As the first choice, since the magnetic strength of real external magnetic black hole should be less than $B_\text{ex}$, we select a random value for the $B_\text{ex}$ case with $0<B<B_\text{ex}$. Contrarily, the GP magnetic strength $B$ is related to GP effect through \eqref{bbr}, so its distribution of $B$ will be less disperse than that of $B_\text{ex}$, by which such GP case can be distinguished from $B_\text{ex}$ dependent case.
 
However, when the error $\delta B$ is large, the dispersion of $B$ distribution due to detector noise should be taken into account. For this purpose, Yuan et. al. \cite{yuan2024,Yuan:2025pbu} proposed the $F$ statistic. Set $\theta_\text{bs}$ as the basis parameter, so here $\theta_\text{bs}$ is magnetic strength $B$ instead of $\rho_0$.
Due to instrumental noise, what we detect for the $i$-th event is not the exact true value of ${\theta_{\text{bs}}}_i$, but a deviated one ${\theta_{\text{bs}}}_i^\prime$ that we can model it as a random value with normal distribution ${\theta_{\text{bs}}}_i^\prime\sim N\left({\theta_{\text{bs}}}_i,\delta{\theta_{\text{bs}}}_i\right)$. With the center values ${\theta_{\text{bs}}}_i^\prime$, the means of them $\overline{{\theta_{\text{bs}}}^\prime}$ 
\begin{equation}
	\overline{{\theta_{\text{bs}}}^\prime}=\frac{1}{n}\sum_{i=1}^n{\theta_{\text{bs}}}_i^\prime.
\end{equation}
and $\delta{\theta_{\text{bs}}}_i$, the statistic $F$ is \textcolor{black}{defined by} \cite{yuan2024,Yuan:2025pbu},
\begin{equation}
	F=\frac{\sum_{i=1}^n\left({\theta_{\text{bs}}}^\prime_i-\overline{{\theta_{\text{bs}}}^\prime}\right)^2}{\sum_{i=1}^n\delta {\theta_{\text{bs}}}_i^2}\label{equn:stat}.
\end{equation} 
The numerator of $F$ is simply the variances for observing ${\theta_{\text{bs}}}^\prime_i$ of considered events, depending both on the source parameter and the detector sensitivity. While the denominator of $F$ is the sum of the measuring precision for the events, reflecting effect of the detector sensitivity. Therefore, if $\theta_{\text{bs}}$ greatly depends on source parameter and the resulting variation much exceeds measuring precision, $F$ will be large. However, when $\theta_{\text{bs}}$ does not quite change with different sources (specially for modified gravity theory effect, $\theta_{\text{bs}}$ is a constant), so that its variation becomes much smaller than the measuring error $\delta\theta_{\text{bs}}$, $F$ would be close to 1 because the numerator and denominator of $F$ approaches the same.

\begin{figure}
	\begin{center}
		\includegraphics[scale=0.6]{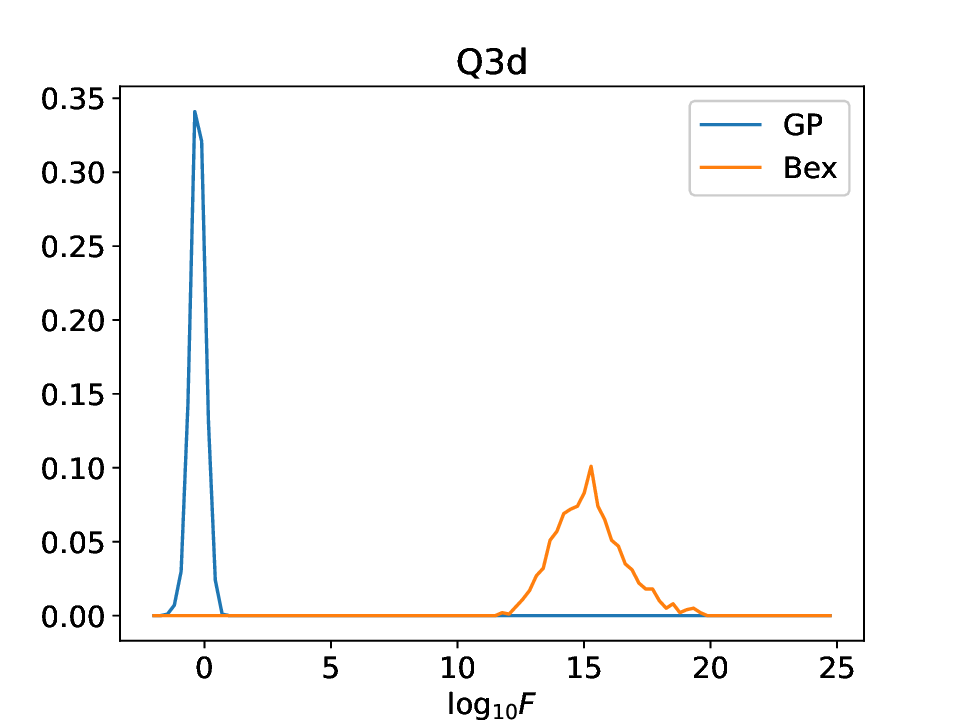}
		\caption{Comparison of the probability distributions of the $F$ statistic for the Q3d source, corresponding to GP and $B_\text{ex}$ dependent cases.}\label{fjhq3dbexsjrhb}
	\end{center}
\end{figure}

\begin{figure}
	\begin{center}
		\includegraphics[scale=0.6]{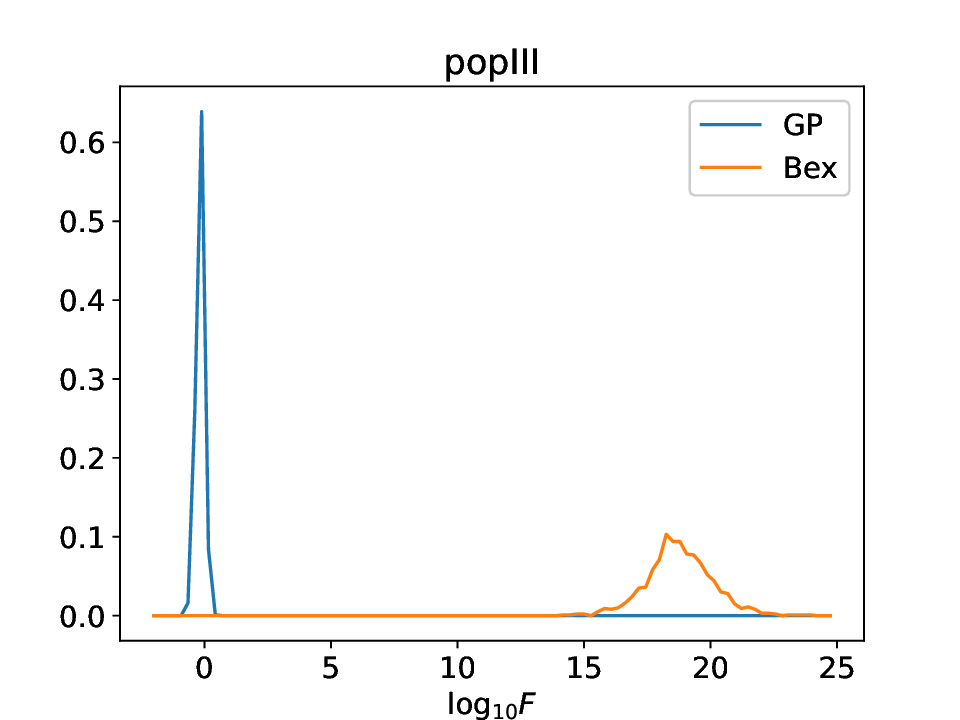}
		\caption{Comparison of the probability distributions of the $F$ statistic for the popIII source, corresponding to GP and $B_\text{ex}$ dependent cases.}\label{fjhq3dbexsjrhbp3}
	\end{center}
\end{figure}

\begin{figure}
	\begin{center}
		\includegraphics[scale=0.6]{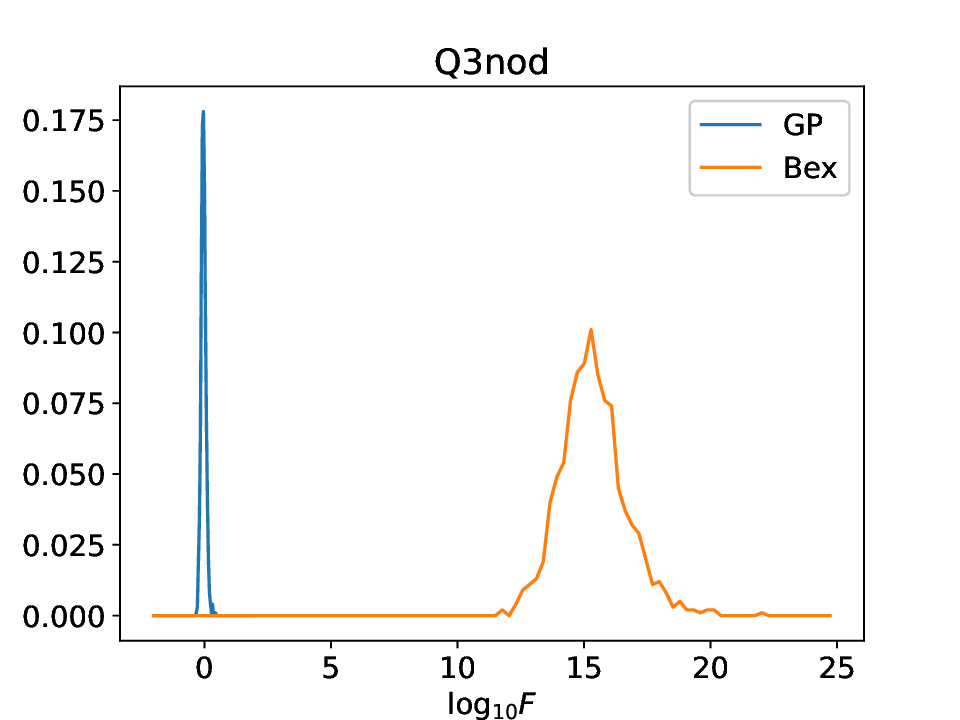}
		\caption{Comparison of the probability distributions of the $F$ statistic for the Q3nod source, corresponding to GP and $B_\text{ex}$ dependent cases.}\label{fjhq3drhbbexsjq3nod}
	\end{center}
\end{figure}

Consider 1000 groups of Q3d sources, the statistic $F$ \eqref{equn:stat} can be calculated for each group, for the $B_\text{ex}$ and GP depending cases, with their $\log_{10}F$ distribution of the 1000 groups shown in \figref{fjhq3dbexsjrhb}. Because the GP case is constrained by the relation \eqref{bbr}, its $\log_{10}F$ is much less than the $B_\text{ex}$ dependent case. The real magnetic effect and GP origin effect thus can be effectively distinguished apart by the statistic $F$: real magnetic effect represented by $B_\text{ex}$ has much larger $\log_{10}F$ than the GP case. The result for PIII and Q3nod illustrated in \figref{fjhq3dbexsjrhbp3} and \figref{fjhq3drhbbexsjq3nod} is similar. Former distinguishment between varying $G$ effect and dynamical friction from dark matter spike in \cite{yuan2024} also obtained such clearcut result.


\subsubsection{Same mean of $\log_{10}B$ as gravitational pull corresponding}\label{smgp}

\begin{figure}
	\begin{center}
		\includegraphics[scale=0.55]{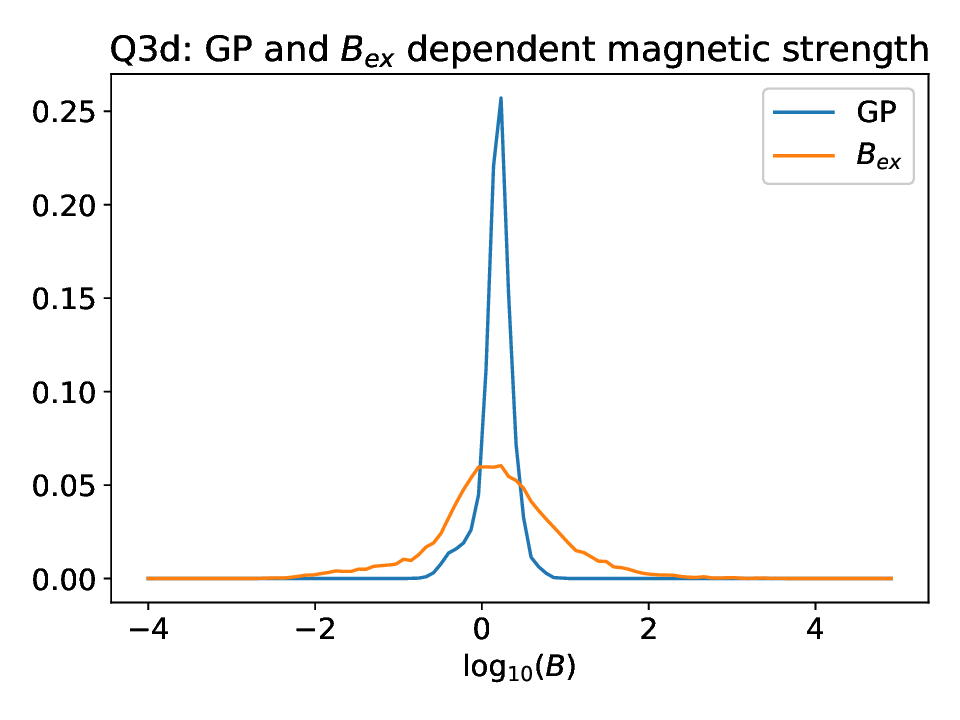}
		\caption{Q3d sources: distribution of ${\theta_{\text{bs}}}_i$ for gravitational pull (GP) and $B_\text{ex}$ dependent magnetic strength $B=kB_\text{ex}$.}\label{tedgdrhbkbrrhbjq3d}
	\end{center}
\end{figure}

\begin{figure}
	\begin{center}
		\includegraphics[scale=0.55]{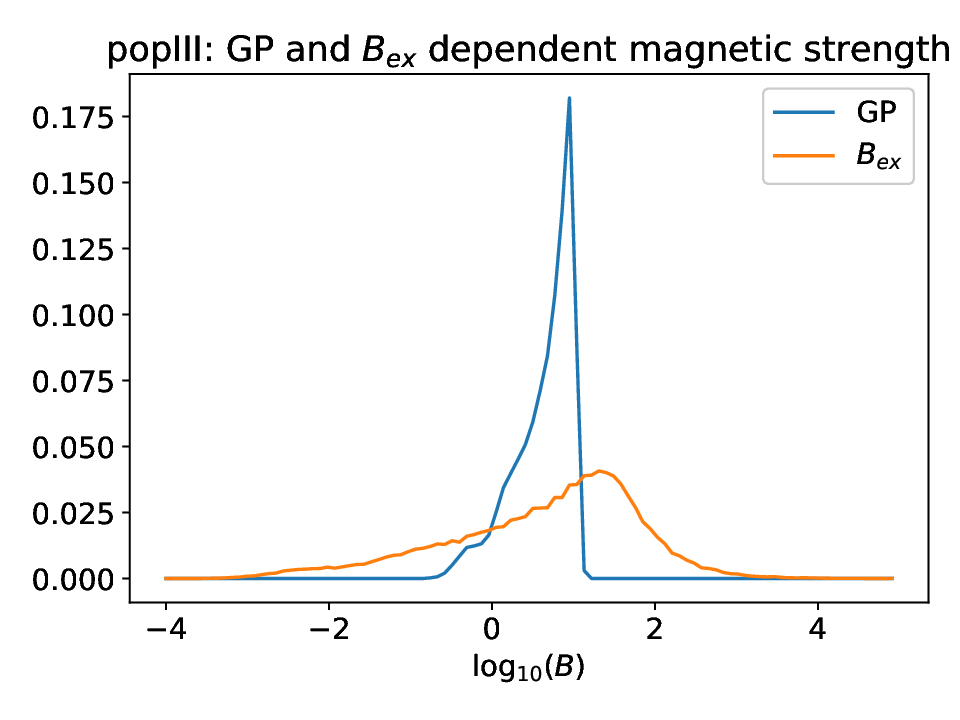}
		\caption{popIII sources: distribution of ${\theta_{\text{bs}}}_i$ gravitational pull (GP) and $B_\text{ex}$ dependent magnetic strength $B=kB_\text{ex}$.}\label{tedgdrhbkbrrhbjp3}
	\end{center}
\end{figure}

\begin{figure}
	\begin{center}
		\includegraphics[scale=0.55]{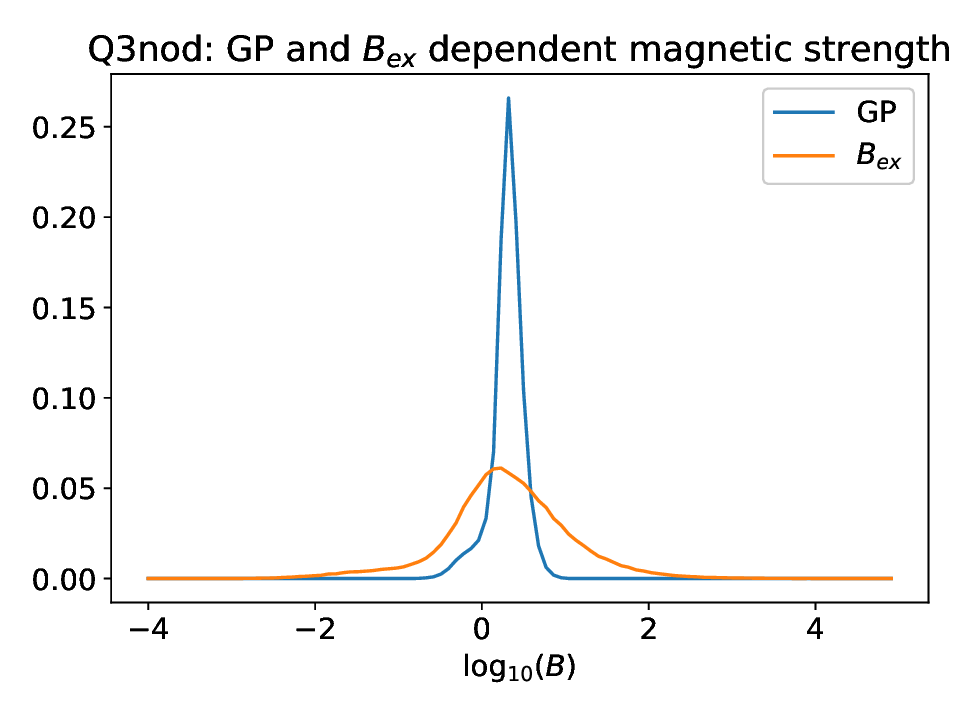}
		\caption{Q3nod sources: distribution of ${\theta_{\text{bs}}}_i$ gravitational pull (GP) and $B_\text{ex}$ dependent magnetic strength $B=kB_\text{ex}$.}\label{tedgdrhbkbrrhbjq3nod}
	\end{center}
\end{figure}

However, besides the condition $0<B<B_\text{ex}$, real magnetic effect may be constrained by some relation which makes it more similar to GP distribution given by \eqref{bbr}, and degrades the distinguishing efficiency. 
As the second choice of $B_\text{ex}$ dependent $B$, we then select the $B_\text{ex}$ dependent magnetic strength as $B=kB_\text{ex}$ such that it has the same mean value of $\log_{10}B$ as the GP dependent distribution. These are displayed in \figref{tedgdrhbkbrrhbjq3d}, \figref{tedgdrhbkbrrhbjp3} and \figref{tedgdrhbkbrrhbjq3nod}.

For a model considered, say Q3d in \figref{tedgdrhbkbrrhbjq3d}, the $B_\text{ex}$ dependent distribution of $B$ looks more disperse, so one can distinguish them easily if the measuring precision $\delta B$ is small enough to probe the profiles of $B$. On the contrary, if $\delta B$ is two large, dependence of $B$ on source parameter will be covered by it. This means the numerator of statistic $F$ will be dominated by measuring error, approaching the same as the denominator. Then although the $B_\text{ex}$ dependent and GP dependent cases have distinct reliance on sources parameter, their numerator of $F$ will both be almost the same as the denominator. Therefore, the distribution of $\log_{10}F$ of real magnetic effect and GP induced effect will be mixed up with each other.

\begin{figure}
	\begin{center}
		\includegraphics[scale=0.6]{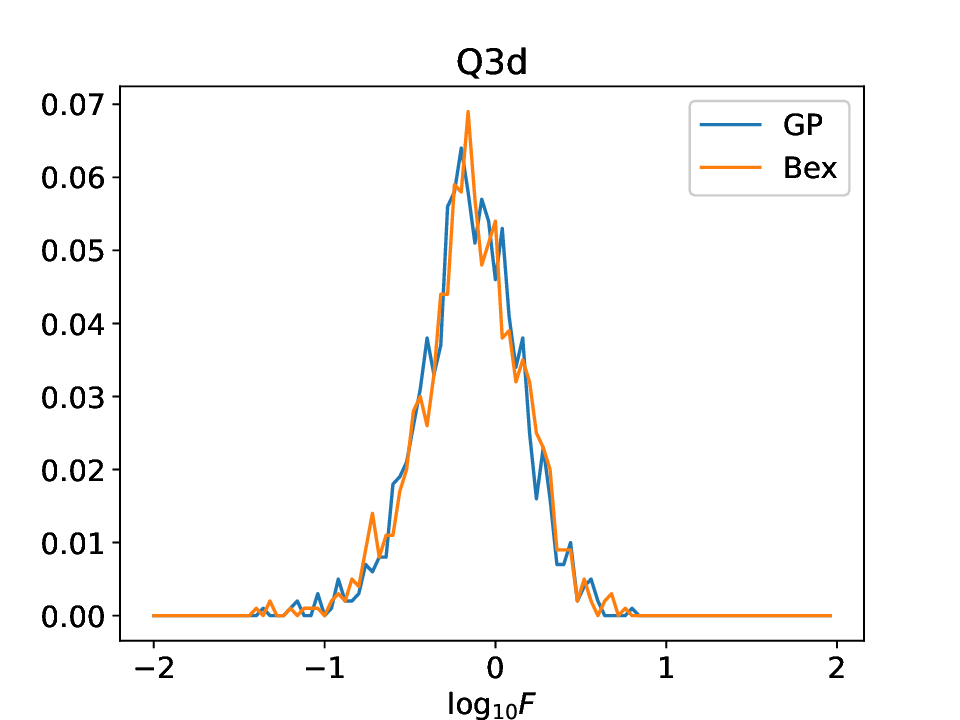}
		\caption{Comparison of the probability distributions of the $F$ statistic for the Q3d source, corresponding to GP and $B_\text{ex}$ dependent cases.}\label{fjhq3dbnrhb}
	\end{center}
\end{figure}

\begin{figure}
	\begin{center}
		\includegraphics[scale=0.6]{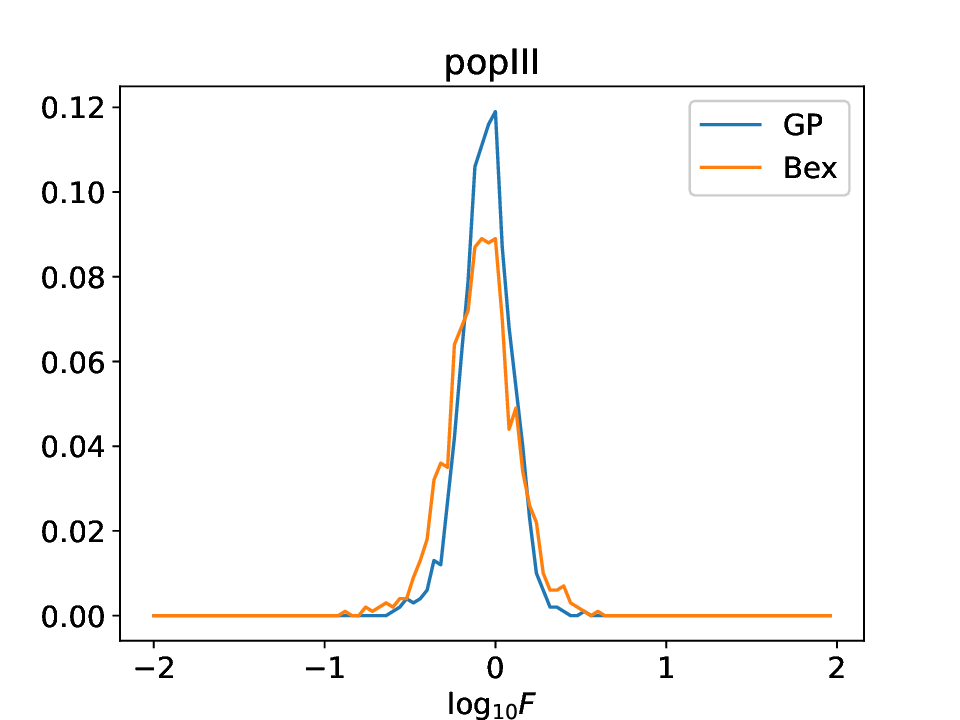}
		\caption{Comparison of the probability distributions of the $F$ statistic for the popIII source, corresponding to GP and $B_\text{ex}$ dependent cases.}\label{fjhq3dbnrhbp3}
	\end{center}
\end{figure}

\begin{figure}
	\begin{center}
		\includegraphics[scale=0.6]{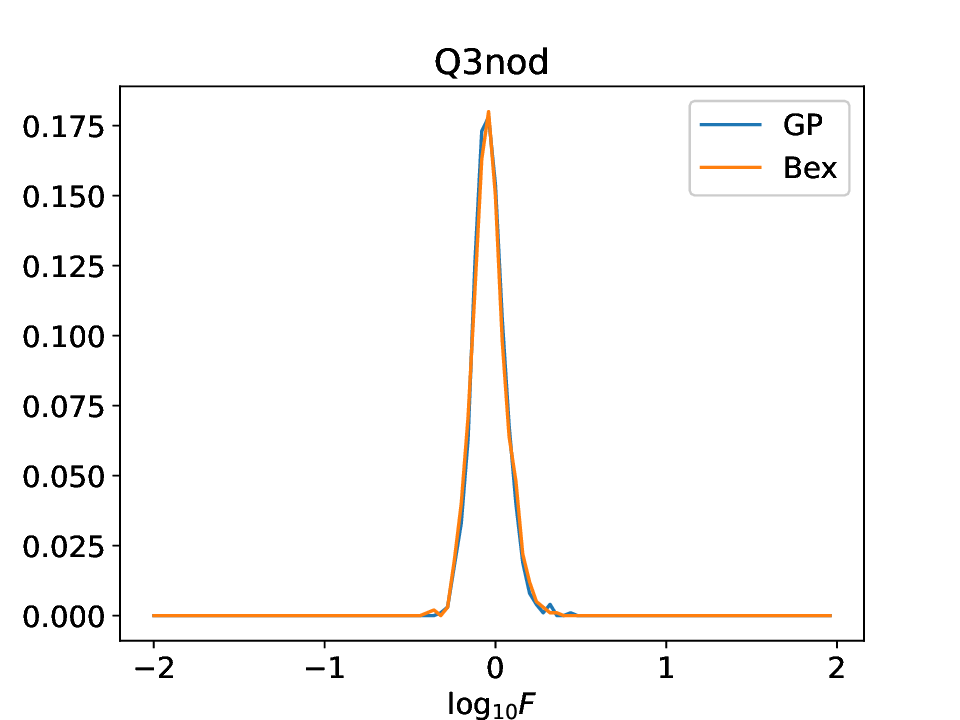}
		\caption{Comparison of the probability distributions of the $F$ statistic for the Q3nod source, corresponding to GP and $B_\text{ex}$ dependent cases.}\label{fjhq3dbnrhbq3nod}
	\end{center}
\end{figure}

For Q3d, the $F$ statistic for each group of source is calculated for matter GP magnetic strength through \eqref{bbr}, and for the $B=kB_\text{ex}$ dependent case respectively, then we plot the distribution of $\log_{10}F$ for the 1000 groups of sources in \figref{fjhq3dbnrhb}. Although the $B$ profile as shown in \figref{tedgdrhbkbrrhbjq3d} for the two cases looks quite different, due to the $\log_{10}B$ that we have set the same on purpose, and that the measuring precision $\delta B$ is very large there, $\log_{10}F$ of the two cases are both very close to $0$, almost indistinguishable. The result for Q3nod sources in \figref{fjhq3dbnrhbq3nod} is similar. Notably popIII has better result, in \figref{fjhq3dbnrhbp3} we see the GP dependent $\log_{10}F$ is more centered at $0$ in contrast to more diverse profile for the $B_\text{ex}$ case, probably thanks to the greater difference of magnetic distribution profile in \figref{tedgdrhbkbrrhbjp3} for the two cases and smaller $\delta B$ for popIII considered.

\begin{figure}
	\begin{center}
		\includegraphics[scale=0.6]{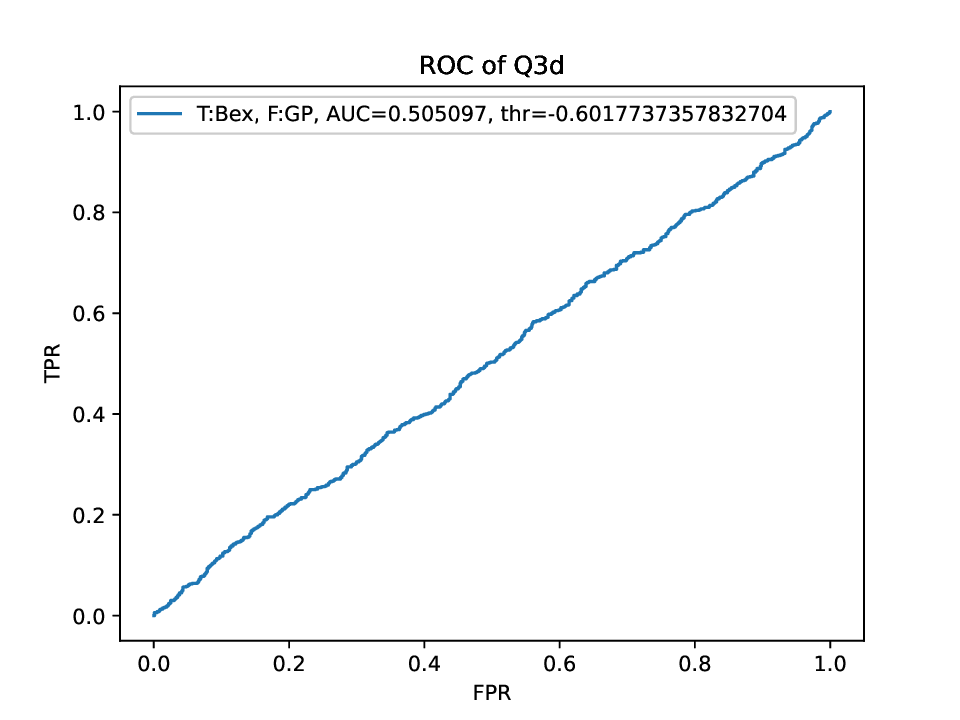}
		\caption{Comparison of the probability distributions of the $F$ statistic for the Q3d source, corresponding to GP and $B_\text{ex}$ dependent cases.}\label{rocrhbbn}
	\end{center}
\end{figure}

\begin{figure}
	\begin{center}
		\includegraphics[scale=0.6]{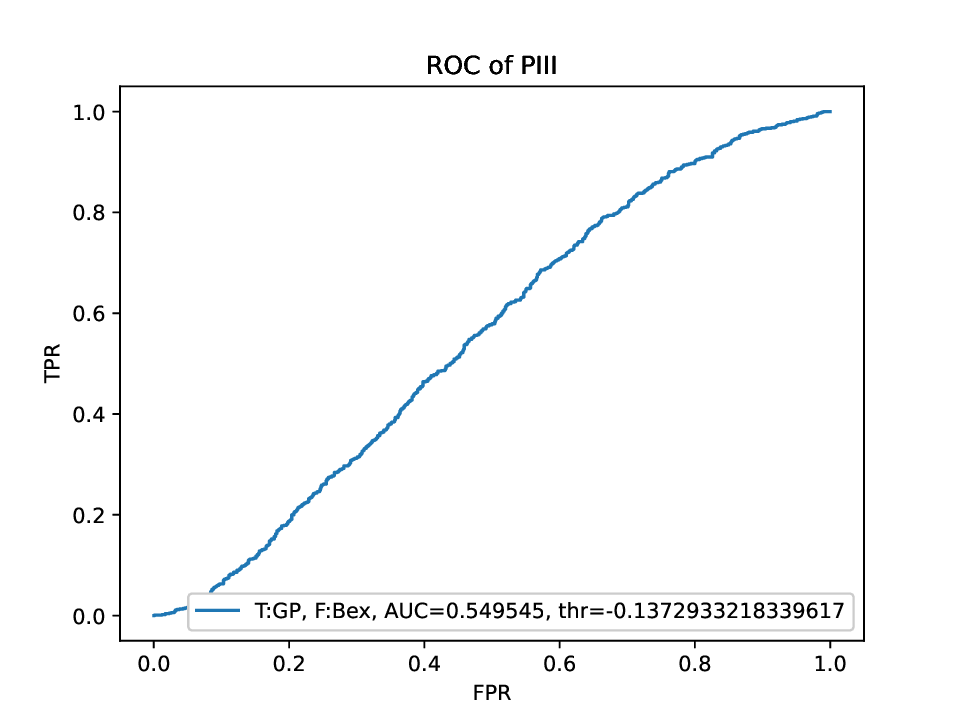}
		\caption{Comparison of the probability distributions of the $F$ statistic for the popIII source, corresponding to GP and $B_\text{ex}$ dependent cases.}\label{rocrhbbnp3}
	\end{center}
\end{figure}

\begin{figure}
	\begin{center}
		\includegraphics[scale=0.6]{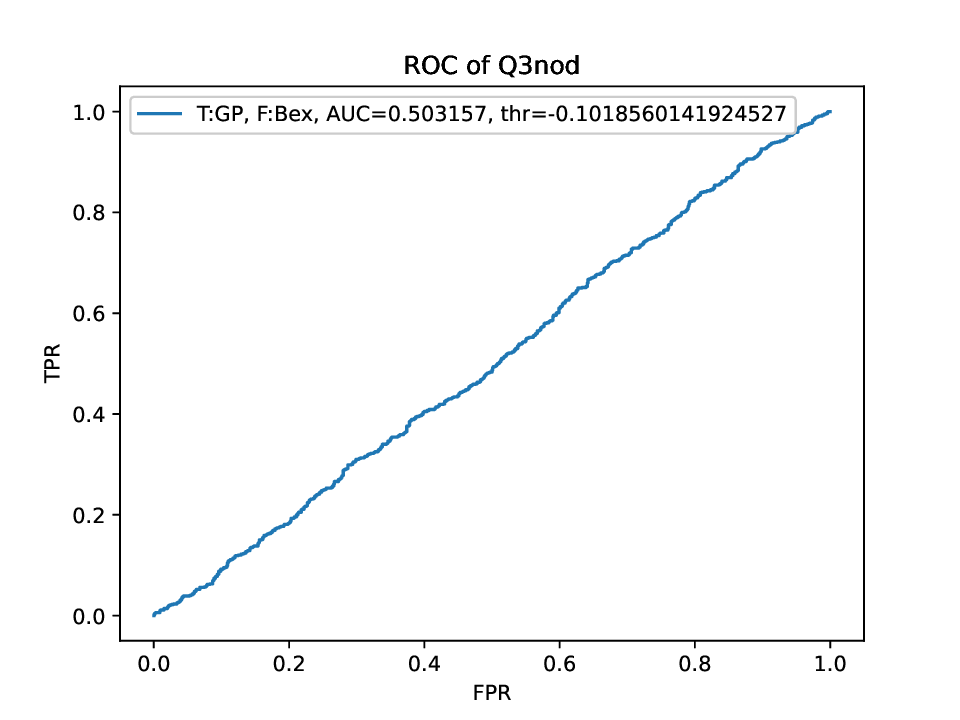}
		\caption{Comparison of the probability distributions of the $F$ statistic for the Q3nod source, corresponding to GP and $B_\text{ex}$ dependent cases.}\label{rocrhbbnq3nod}
	\end{center}
\end{figure}

It is therefore harder to find the distinguishing threshold of statistic $F$ between real magnetic effect and GP corresponding effect, adopting the second choice of $B_\text{ex}$ dependent $B$, compared with the disparate results in \ref{rdbex}.
So next we employ the \ac{ROC} curve method to quantifying these results by investigating how different thresholds influences distinguishing the two effects \cite{Yuan:2025pbu}. First set one of the two cases as positive and the other negative type. Then choose a threshold of $\log_{10}F$ to determine which type the data is classified as, and there is the true positive rate (TPR) and false positive rate (FPR). With changing the threshold, a curve of FPR versus TPR: the \ac{ROC} curve is drawn. The \ac{ROC} curve comparing the gravitational pull from matter and $B_\text{ex}$ dependent cases are plotted in \figref{rocrhbbn}, \figref{rocrhbbnp3} and \figref{rocrhbbnq3nod}. For Q3d and Q3nod models the two effects are nearly indistinguishable with AUC very close to 0.5, consistent with former discussion. And popIII has better result, its AUC is about 0.55. For this model the best distinguishing threshold of $\log_{10}F=-0.1373$ is given by Youden index, the maximum of TPR+TNR-1 which corresponds the the point of \ac{ROC} curve that most approaches the upper-left direction.

To summarize, to distinguish between real external BR magnetic field effect and gravitational pull inducing one, if we adopt the real $B$ as random value below $B_\text{ex}$, we can completely differentiate these two effects with the statistic $F$. Alternatively, if real magnetic field effect is very similar to gravitational pull inducing one, for example we set it as $B=kB_\text{ex}$ with the same $\log_{10}B$ as gravitational pull, they would be nearly indistinguishable, and in this case, $-2$ PN order \ac{GW} corrections from real BR magnetic effect in future observation would be mixed up with those from gravitational pull origin, even considering multiple events with the statistic $F$. And popIII source has slightly better results.

\subsection{Distinguish gravitational pull and Bonnor-Melvin external megnetic field}
\begin{figure}
	\begin{center}
		\includegraphics[scale=0.6]{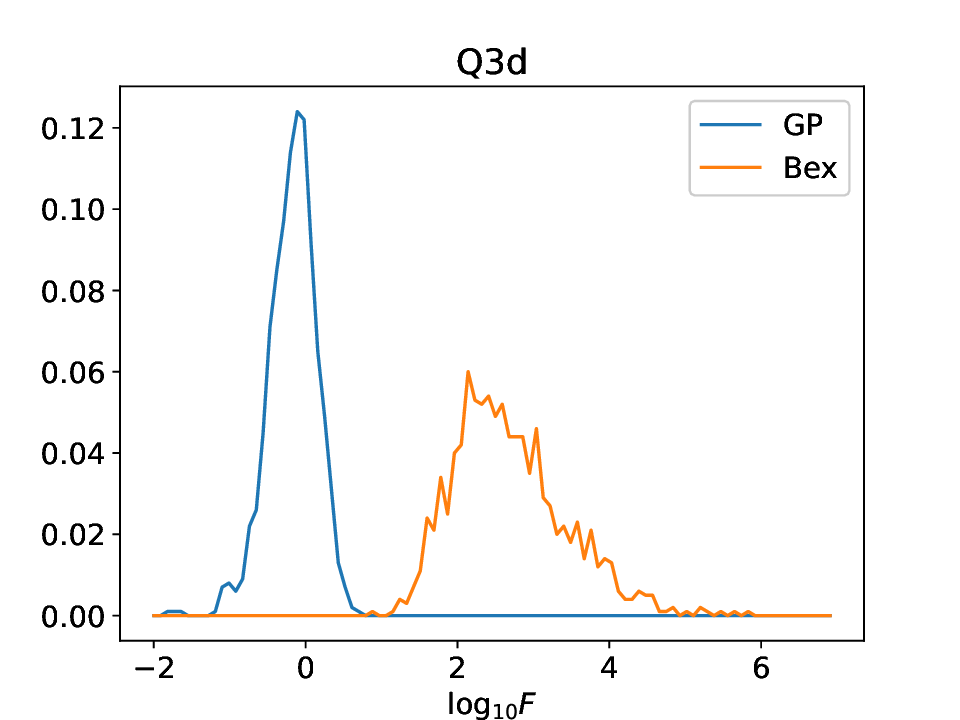}
		\caption{Bonnor-Melvin magnetic effect: probability distributions of the $F$ statistic for the Q3d source, corresponding to GP and $B_\text{ex}$ dependent cases.}\label{fjhq3dbnrhbkbm1e24}
	\end{center}
\end{figure}

\begin{figure}
	\begin{center}
		\includegraphics[scale=0.6]{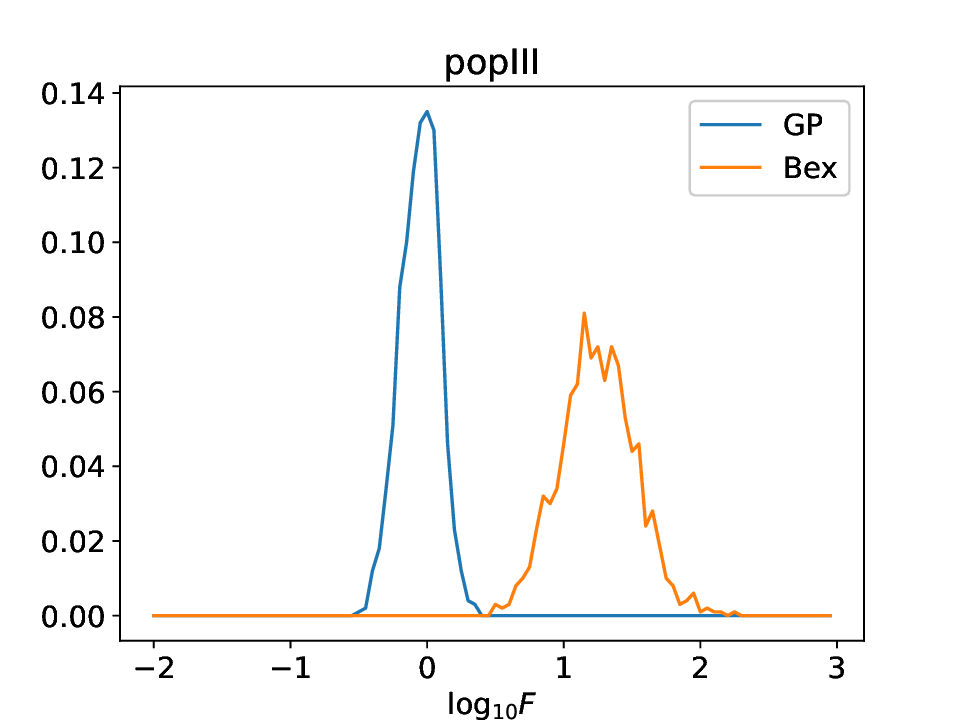}
		\caption{Bonnor-Melvin magnetic effect: distributions of the $F$ statistic for the popIII source, corresponding to GP and $B_\text{ex}$ dependent cases.}\label{fjhq3dbnrhbkbm1e24p3}
	\end{center}
\end{figure}

\begin{figure}
	\begin{center}
		\includegraphics[scale=0.6]{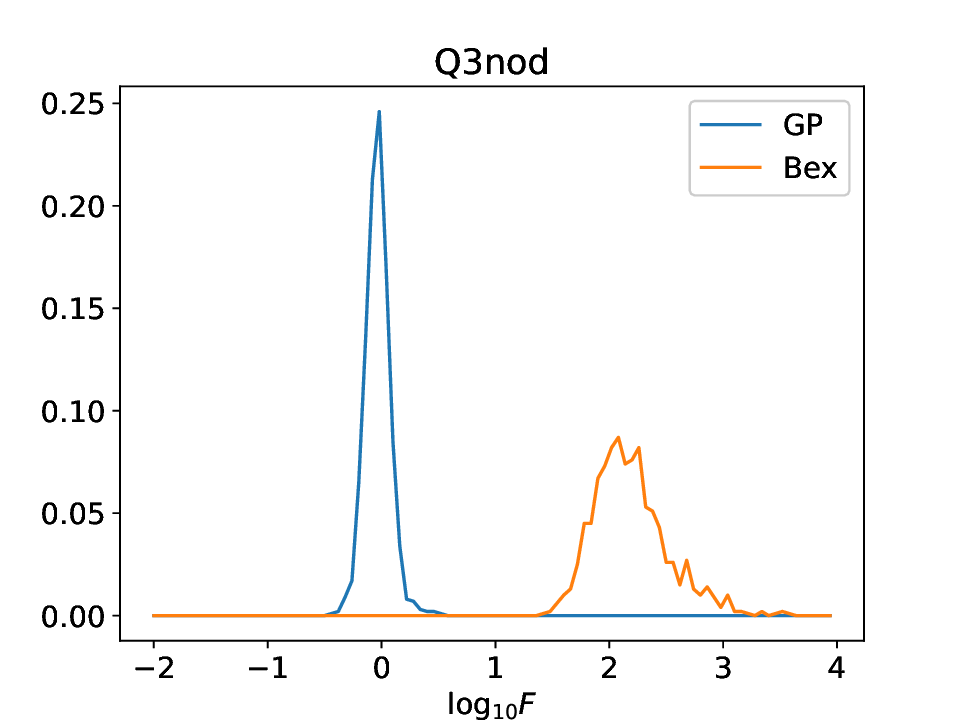}
		\caption{Bonnor-Melvin magnetic effect: distributions of the $F$ statistic for the Q3nod source, corresponding to GP and $B_\text{ex}$ dependent cases.}\label{fjhq3dbnrhbkbm1e24q3nod}
	\end{center}
\end{figure}

Now let's consider another external magnetic effect, the Bonnor-Melvin one, and similar distinguishment can be done as above with the statistic $F$. Real magneitc field effect which we assume depending on $B_\text{ex}$ \eqref{bexkm} are different from that of gravitational pull origin through \eqref{bbm}. With the first choice of real magnetic effect $0<B<B_\text{ex}$, these two cases can be totally distinguished apart like former Bertotti-Robinson result in \figref{fjhq3dbexsjrhb}, \figref{fjhq3dbexsjrhbp3} and \figref{fjhq3drhbbexsjq3nod}, because real magnetic field is randomly distributed so that it's possible to have smaller $\delta B$, while GP corresponding one is more constrained at a weak strength where $\delta B$ is relatively large. However, if real magnetic distribution is quite similar to that from GP, as the second choice, $B=kB_\text{ex}$ with the same $\log_{10}B$ as GP, they can be nearly indistinguishable like Bertotti-Robinson in \figref{fjhq3dbnrhb}, \figref{fjhq3dbnrhbp3} and \figref{fjhq3dbnrhbq3nod}. We have verified this for GP corresponding $B$ \eqref{bbm} with $\rho_0<7.3\times 10^{-7}\text{kg}/\text{m}^3$. Interestingly, when we raise $\rho_0$ to $7.3\times 10^{-1}\text{kg}/\text{m}^3$, the $F$ statistic of real magnetic and GP corresponding case will get significantly distinguishable as shown in \figref{fjhq3dbnrhbkbm1e24}, \figref{fjhq3dbnrhbkbm1e24p3} and \figref{fjhq3dbnrhbkbm1e24q3nod}, even though we have set the two cases with the same $\log_{10}B$. So different from Bertotti-Robinson magnetic effect, there exists a turning value of $\rho_0$ between $7.3\times 10^{-1}\text{kg}/\text{m}^3$ and $7.3\times 10^{-7}\text{kg}/\text{m}^3$, above which real Bonnor-Melvin magnetic effect can be distinguished from gravitational pull origin. Next we are going to find the exact turning value of $\rho_0$, which is estimated around  $10^{-4}\text{kg}/\text{m}^3$ and corresponding magnetic strength $B\sim 10^{4}\text{T}$ according to \eqref{bbm}, where transition of distinguishability between real magnetic effect and gravitational pull occurs.

\section{Conclusions}\label{concl}

In this paper, we derive the \ac{ppE} inspiral waveform corrections for binary black hole systems where one component is a \ac{KBR} or \ac{KBM} —-i.e., a rotating black hole immersed in an external magnetic field. By modifying the orbital energy and Kepler’s law, we obtain the frequency evolution and thereby compute the approximate phase correction within the stationary phase approximation. The leading-order magnetic field corrections for \ac{KBR} and \ac{KBM} black holes appear at $-2$ and $-3$ PN order, respectively. These corrections are not degenerate with those from any standard modified gravity theories, yet they coincide in PN order with the gravitational effects induced by a power-law matter distribution with indices $\gamma_\text{BR}=1$ and $\gamma_\text{BM}=0$. The leading-order spin corrections for both \ac{KBR} and \ac{KBM} black holes are found at $-1.5$ PN order. 

Incorporating higher-order waveform corrections and multiple spherical harmonic modes, we assess Tianqin’s capability to constrain the magnetic field strength through the inspiral \ac{GW} signals of massive binary black holes. Our analysis shows that systems with lower chirp mass and larger symmetric mass ratio $\eta$ yield better measurement precision $\delta B$, and the relative precision gets better for smaller $\eta$. Under three astrophysical source models Q3d, PIII and Q3nod, —the light-seed model provides the most stringent constraint, achieving a precision down to $\delta B$ as low as $10^{-2}\text{T}$.

Given that the leading magnetic corrections enter at $-2$ and $-3$ PN orders, they can be observationally degenerate with the \textcolor{black}{gravitational pull} 
of a power-law matter distribution characterized by $\gamma_\text{BR}=1$ and $\gamma_\tx{BM}=0$. We thus establish a mapping between the magnetic field strength $B$ and the matter density $\rho_0$ that produces equivalent waveform distortions. Consequently, a future detection of a $-2$ or $-3$ PN correction in the \ac{GW} signal could be attributed to either ambient matter or the intrinsic magnetic field of \ac{KBR}/\ac{KBM} black holes.

To distinguish between these two physical origins, \textcolor{black}{we then} apply the model-selection procedure proposed by Yuan et al.~\cite{yuan2024,Yuan:2025pbu} and calculate the statistic $F$ for each 1000 groups of Q3d, popIII or Q3nod sources to identify real KBR magnetic effect and effect mimicked by gravitational pull. We model the real magnetic strength as one related to the extremal value $B_\text{ex}$ of black hole in some way, compared with the gravitational pull corresponding one that depends on matter density $\rho_0$ but actually does not vary much for all considered sources. We first set the $B_\text{ex}$ dependent $B$ as a random value below $B_\text{ex}$, then the resulting $F$ statistic is much greater than that of the gravitational pull corresponding case, so real magnetic effect ($\log_{10}F\approx0$) can be completely distinguished from gravitational pull effect ($\log_{10}F>10$). However, if real magnetic effect is enough similar to gravitational pull where we assume real $B$ as $B=kB_\text{ex}$ and has the same mean of $\log_{10}B$ as gravitational pull case, the statistic $F$ of the two cases would be highly overlapped ( both $\log_{10}F\approx0$), and real magnetic effect is actually nearly indistinguishable from gravitational pull, even though considering multiple events. The reason, besides the highly overlapped of $\log_{10}B$, includes large precision $\delta B$ at such low magnetic strength.  Only popIII sources exhibit slightly distinguishability.

The distinguishment considering Bonnor-Melvin magnetic effect is similar but more interesting. Even when real magnetic effect is very similar to gravitational pull corresponding one where we adopt the second choice of real $B$ as $B=kB_\text{ex}$ with the same mean of $\log_{10}B$ as gravitational pull case, their $F$ statistic distribution can be distinct, and real magnetic effect can be distinguished from GP as long as the $\rho_0$ is high enough. We estimate the transition value of $\rho_0$ as $10^{-4}\text{kg}/\text{m}^3$ and corresponding $B\sim 10^{4}\text{T}$, thus next we are going to find how distinguishability behaves between this interval, for example, implement threshold selection with ROC curve and see how AUC transits from 1 to 0.5.

Therefore, due to the degeneracy between magnetic fields and environmental effects, it is highly worthwhile to distinguish them through multi-messenger observations or validation across multiple events in the future when they are not strong enough and too similar. Although Bertotti-Robinson external magnetic effect can be probably too similar to gravitational pull to be distinguished from the latter even with multiple events, Bonnor-Melvin magnetic effect has more chance to do it. 

\section*{Acknowledgements}
This work is supported by National Natural Science Foundation of China (NSFC) with Grant No. 12275087

\bibliography{references}

\end{document}